\documentclass[10pt,aps,prd,twocolumn,a4paper,showpacs,nofootinbib]{revtex4-1}
\usepackage[utf8x]{inputenc}
\usepackage{amsmath}
\usepackage{amsfonts}
\usepackage[colorlinks=true,linkcolor=blue,citecolor=blue,urlcolor=blue]{hyperref}
\usepackage{amssymb}
\usepackage{graphicx}
\usepackage{csquotes}
\usepackage[rightcaption]{sidecap}
\sidecaptionvpos{figure}{c}

\renewcommand{\Re}{\textrm{Re}}

\newcommand{\bk}{\mathbf{k}}
\newcommand{\eq}[1]{Eq.~\eqref{#1}}
\providecommand{\abs}[1]{\left|#1\right|}
\providecommand{\ep}[1]{{e}^{#1}}
\newcommand{\Tr}{{\rm Tr}}
\newcommand{\PH}{Peres-Horodecki }

\begin{document}
\title{Quantum entanglement in analogue Hawking radiation, \texorpdfstring{\\}{ }
when is the final state non-separable ?} 

\author{Xavier Busch}\email[]{xavier.busch@th.u-psud.fr}
\author{Renaud Parentani}
\email[]{renaud.parentani@th.u-psud.fr}
\affiliation{Laboratoire de Physique Th\'eorique, CNRS UMR 8627, B{\^{a}}timent\ 210,
         \\Universit\'e Paris-Sud 11, 91405 Orsay CEDEX, France}
         

\begin{abstract}
We study the quantum entanglement of the quasiparticle pairs emitted by analogue black holes. We use a phenomenological description of the spectra in dispersive media to study the domains in parameter space where the final state is non-separable. In stationary flows, three modes are involved in each sector of fixed frequency, and not two as in homogeneous situations. The third spectator mode acts as an environment for the pairs, and the strength of the coupling significantly reduces the quantum coherence. The non-separability of the pairs emitted by white holes are also considered, and compared with that of black holes. 
\end{abstract}

\maketitle

\section{Introduction}

One of the main challenges of the analogue gravity program is to conceive and realize experiments where a clear signal of the analogue Hawking effect would be detected~\cite{Unruh:1980cg,Barcelo:2005fc}. When addressing this question, one should clearly distinguish the induced effect, which purely rests on the dynamics of classical fields, i.e., the scattering of incident waves~\cite{Schutzhold:2002rf,Rousseaux:2007is,Weinfurtner:2010nu}, from the spontaneous effect which arises from the amplification of vacuum fluctuations~\cite{Unruh:1994je,Brout:1995wp}. However, because the same mode amplification is involved, both channels lead to very similar behaviors. Indeed, the space-time properties of the correlation patterns of the emitted quasi-particles are very much the same whether or not the spontaneous channel significantly contributed~\cite{Carusotto:2008ep,Macher:2009nz,Recati:2009ya}. Therefore, if one wishes to experimentally distinguish the spontaneous from the induced, one must use observables that are sensitive to the small differences between the quantum and the classical.

The same question arises in a simpler context, namely pair creation caused by a temporal change in an homogeneous medium~\cite{Bruschi:2013tza,Busch:2013sma,Busch:2013gna,PhysRevLett.108.260401,WilsonDCE}. Because of the homogeneity, the state of linear perturbations splits into two-mode sectors characterized by opposite wave vectors $\bk$. Therefore, the entanglement between $\bk$ and $-\bk$ can be analyzed separately. To distinguish quantum states from classical distributions, it suffices to compare the strength of the correlations between $\pm \bk $, which is given by the norm of $c_k = \Tr{[\hat \rho \, \hat a_{-\bk} \hat a_\bk]}$, with the mean occupation numbers $n_{\bk} = \Tr{[\hat \rho \, \hat a^\dagger_\bk \hat a_\bk]}$, where $\hat a^\dagger_\bk$ ($\hat a_\bk$) corresponds to the creation (destruction) operator of a quasi-particle of wave number $\bk$. In fact whenever 
\begin{equation}
\begin{split}
\Delta_k \doteq n_\bk \, n_{-\bk} - \abs{c_k}^2 < 0,
\label{cnk}
\end{split}
\end{equation}
the two-mode system is quantum mechanically entangled, because classical correlations obey $\Delta_k \geq 0$.\footnote{
A brief account of these concepts, as well as their relationship with the more elaborate \PH criterion is given in Appendix~\ref{app:separability}. Notice also that, as a matter of principles, \eq{cnk} equally applies both to cosmological pair creation in the early universe~\cite{Campo:2005sy,Campo:2005sv,Campo:2008ju}, and to analogue situations~\cite{PhysRevLett.108.260401,WilsonDCE,Bruschi:2013tza,Busch:2013sma,Busch:2013gna}.} 
Hence, when the initial state contains no correlations between $\pm\bk$, a violation of \eq{cnk} by the final values of $c_k$ and $n_\bk$ implies that the spontaneous channel significantly contributed. 

When considering inhomogeneous and stationary flows containing one sonic horizon, the situation is similar in that stationary states factorize into independent sectors of fixed frequency $|\omega|$. Yet the situation is more complex because, for low frequency, each sector contains three modes. Two have positive norm, and describe co- and counter propagating quasiparticles with respect to the fluid. The third mode has a negative norm, and describes the negative energy partners trapped in the supersonic region. Two types of entangled pairs can thus be created. In spite of the fact that the modes interact with each other, the criterion of \eq{cnk} with $\bk \to \omega$ applies to each pair considered separately. In fact, the leftover third mode acts as an environment for the considered pair, thereby giving a situation similar to that studied in~\cite{Busch:2013sma,Busch:2013gna}. 

Following these works, we aim to characterize the domains in parameter space where the final state is non-separable. To this end, we first select a set of relevant parameters which span this space. We then adopt a phenomenological description of the spectra in dispersive media. Combining these, our analysis reveals the crucial roles played by the strength of the (two) coupling parameters between the leftover mode and the two-mode system under study. The present analysis overlaps with Refs.~\cite{deNova:2012hm,finazzi2013} where similar issues are considered. It complements these works in several respects: First, we parametrize a much wider class of situations, and we show that the entanglement follows two scaling laws which are associated to the two regimes found in the spectral analysis of Ref.~\cite{Finazzi:2012iu}. Second, we compare the non-separability of the pairs emitted by black and white hole flows. 
 
The paper is organized as follows. In Section \ref{sec:system}, we parametrize the coefficients of the $S$-matrix on a black hole horizon, and we identify the relevant set of parameters which govern the final state of the quasi-particles. In Section \ref{sec:domainsofnonsep}, we study the domains of parameter space where the final state is non-separable. In Appendix \ref{app:separability}, we review the key concepts governing the notion of non-separability. In Appendix \ref{app:sublum}, we briefly compare the strength of the entanglement when replacing the superluminal dispersion used in the body by a subluminal one. In Appendix \ref{app:whitehole}, we study the entanglement of pairs emitted by white hole flows. 

\section{The system}
\label{sec:system}

We consider a dispersive quantum field $\hat \phi$ describing linear perturbations propagating in a one dimensional moving medium. The flow is stationary and characterized by $v(x)$. For simplicity we assume that the speed of low frequency waves is constant and taken to be unity. In this case, the field equation is~\cite{Unruh:1994je}
\begin{equation}
\begin{split}
\left [(\partial_t - \partial_x v) (\partial_t - v \partial_x) - F^2(i\partial_x)\right ] \hat 
\phi(t,x) = 0 ,
\label{waveeq}
\end{split}
\end{equation}
where $F^2(k)$ gives the dispersion relation. In the body of the text, we consider the superluminal law
\begin{equation}
\label{eq:disprelsuper}
\begin{split}
F^2(k) &= k^2 ( 1+ \frac{k^2}{\Lambda^2}) ,
\end{split}
\end{equation}
where the wave number $\Lambda$ fixes the dispersive scale. The quartic subluminal case is treated in App.~\ref{app:sublum}.

Because the flow is stationary, the solution of \eq{waveeq} splits into $\omega$ sectors: $\hat \phi_\omega(x) = \int dt e^{- i \omega t} \hat \phi(t,x)$ which can be studied separately~\cite{Brout:1995wp}. When the flow is asymptotically uniform on both sides, the incoming modes $\phi^{\rm in, a}_\omega$, with a single branch with group velocity pointing towards the horizon, are well defined and asymptotically superpositions of plane waves. (The index $a$ refers to the dimensionality of the set of solutions at fixed $|\omega|$.) The same apply to the outgoing modes $\phi_\omega^{\rm out, a}$, with the group velocity pointing now away from the horizon. More details about this identification and the dimensionality of the set of modes can be found in \cite{Macher:2009nz,Macher:2009tw}. 

\begin{figure}
\includegraphics[width=1\linewidth]{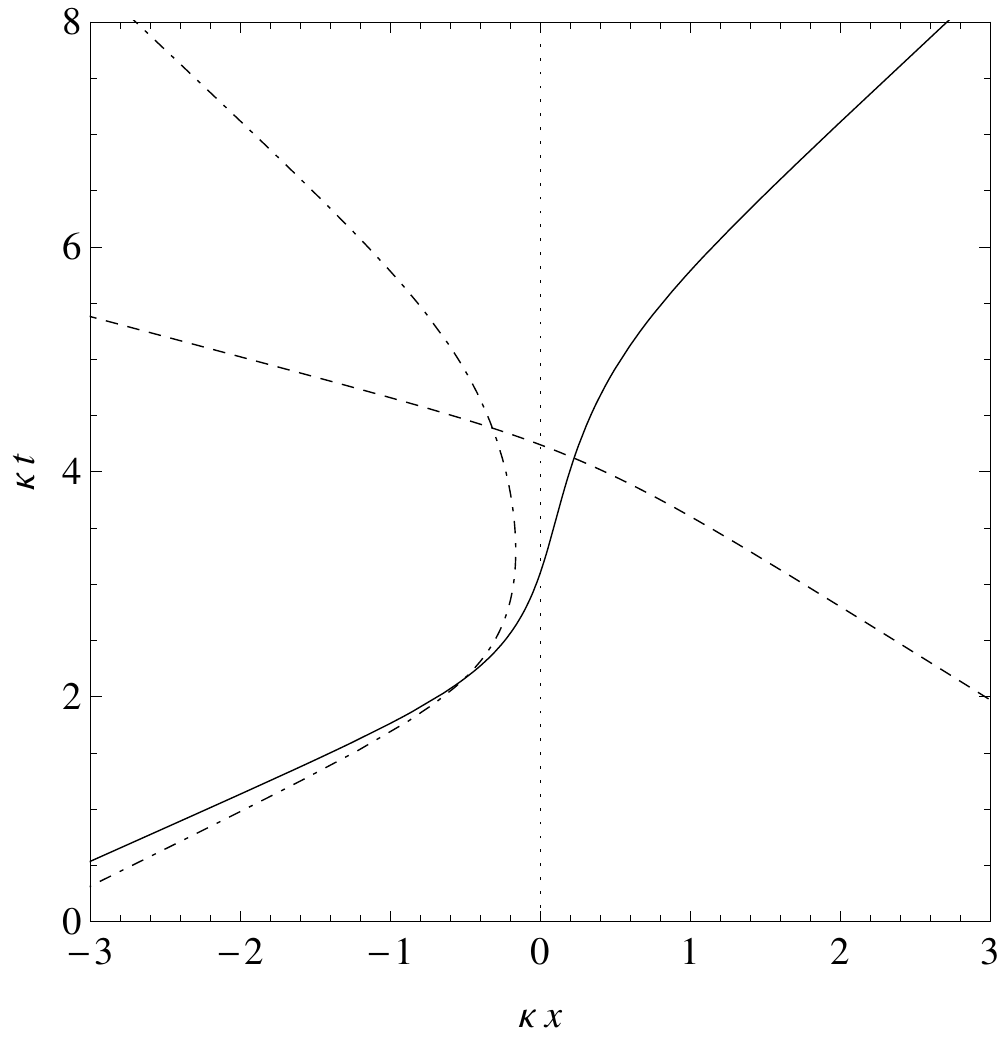}
\caption{We represent the space-time characteristics associated with the three dispersive modes for Eq.~\eqref{eq:disprelsuper} and in a black hole flow with $v < 0$, see Eq.~\eqref{eq:vpluscBH} for an example. The horizon $v = - 1$ is located at $x = 0$, and represented by the vertical dotted line, as the time runs vertically. On the right (left) the flow is sub (super) sonic. The solid (dot-dashed) line ending in the upper right (left) corner describes the $u$ mode with positive (negative) frequency which escapes (is trapped). The dispersive effects are clearly visible in the past. The dashed line represents the co-propagating $v$ mode which crosses the horizon with no significant change. }
\label{fig:caracsuper}
\end{figure}

In brief, because of superluminal dispersion, there is a threshold value $\omega_{\rm max}$ above which there is no pair creation. For $0< \omega< \omega_{\rm max}$, there are three independent modes. They are called $\phi_{\omega}^{\rm in, u}, \,\phi_{\omega}^{\rm in, v}, (\phi_{-\omega}^{\rm in, u})^*$ for the $in$ modes (and similarly for the $out$ ones). The first two have positive norm and describe respectively counter- and copropagating quasi-particles, see Fig.~\ref{fig:caracsuper}. The third one, $(\phi_{-\omega}^{\rm in, u})^*$, has a negative norm, and describes the incoming negative frequency partner trapped in the supersonic region. These three modes are scattered in the near horizon region. As a result, the outgoing modes are non trivially related to the incident ones. The $S$-matrix relating the (normalized) $in$ modes to the $out$ ones is thus an element of $U(1,2)$. Following~\cite{Macher:2009nz}, we name its coefficients 
\begin{equation}
\begin{split}
\left ( \begin{array}{l}
\hat a_\omega^{u } \\
(\hat a_{-\omega}^{u} )^\dagger\\
\hat a_\omega^{v} 
\end{array} \right ) = \left ( \begin{array}{lll}
\alpha_\omega & \beta_\omega^*& A_\omega \\
\beta_{-\omega}& \alpha_{-\omega}^* & B_\omega \\
\tilde A_\omega & \tilde B_\omega^* & \alpha^v_\omega 
\end{array} \right ) \left ( \begin{array}{l}
\hat a_\omega^{u,\rm in}\\
  (\hat a_{-\omega}^{u,\rm in})^\dagger \\
 \hat a_\omega^{v,\rm in}
\end{array}\right ).
\label{Bogcoef}
\end{split}
\end{equation}
To avoid ponderous notation, only the superscript ``in'' will be written. The superscript ``out'' is thus implied. The 3 independent pairs of destruction and creation operators associated with the 3 $in$ (or 3 $out$) modes obey the canonical commutation relations. 
 
When $\hat \rho$, the state of the quantum field, is stationary and Gaussian, it factorizes into 3-mode sectors of fixed $|\omega|$. In this paper we shall only consider such states. Then because the $S$-matrix of \eq{Bogcoef} only mixes modes with the same $|\omega|$, the factorization equally applies to the description of $\hat \rho$ in terms of $in$, or $out$ quasiparticle content. We shall study the latter since we aim to identify the cases where the state after the scattering is non-separable. 

For $\omega >0$ and at late time, each 3-mode state is fully characterized by six numbers 
\begin{equation}
\begin{split}
\label{eq:noutcoutdef}
n^{u}_{\pm\omega} &= {\rm Tr} \left ( \hat \rho \, (\hat a_{\pm\omega}^{u})^\dagger \hat a_{\pm\omega}^{u}\right ),\quad  n^v_\omega = {\rm Tr} \left ( \hat \rho \, (\hat a_\omega^{v})^\dagger \hat a_\omega^{v}\right ),\\
c^{u u/v}_\omega &= {\rm Tr} \left ( \hat \rho \, \hat a_{-\omega}^{u} \hat a_\omega^{u/v}\right ),\quad\ \, 
d^{uv}_\omega = {\rm Tr} \left ( \hat \rho \, (\hat a_\omega^{u})^\dagger \hat a_\omega^{v}\right ).
\end{split}
\end{equation}
The interpretation of the 3 (real and positive) final occupation numbers $n^a_\omega$ is straightforward and standard. The two $c_\omega$ are complex, and their norm quantify the strength of the statistical correlations between outgoing quasi-particles of opposite energy, namely between the $uu$ pairs of counter-propagating $out$ modes $(\phi_{\omega}^{\rm out, u}, \phi_{-\omega}^{\rm out, u})$, and the $uv$ pairs $(\phi_{\omega}^{\rm out, v}, \phi_{-\omega}^{\rm out, u})$. They generalize the $c_k$ term which enters into \eq{cnk}. In the present case, we thus have two differences
\begin{subequations}
\label{eq:NSepuv}
\begin{align}
\Delta_\omega^{uu}  \doteq  n_{-\omega}^{u} n_\omega^{u} - \abs{c_\omega^{uu}}^2 , \\
\Delta_\omega^{uv}  \doteq  n_{-\omega}^{u} n_\omega^{v} - \abs{c_\omega^{uv}}^2 .
\end{align}
\end{subequations}
We show in appendix~\ref{app:separability} that if one of them is negative, the state is non separable. The last coefficient of \eq{eq:noutcoutdef}, $d^{uv}_{\omega}$, characterizes the strength of the correlations between $u$ and $v$ modes which have been elastically scattered. Thus it results from the analogue ``grey body'' factors. These correlations are never strong enough to violate classical inequalities, see Eq.~\eqref{eq:valuePHcritcis0}. Hence, they shall no longer be mentioned. 

In order to be able to determine if the final state is non-separable, it is necessary to know the initial state and the coefficients of the $S$-matrix. Our aim is not so much to perform the calculation in a particular realization, rather we aim to characterize the domains in parameter space where the state is non-separable. To this end, we need to identify the independent parameters which span this space, and to adopt a phenomenological description of their behaviors.

As a first step, we assume that the initial state is incoherent. It is thus characterized by the 3 initial occupations numbers $n^{\rm in, u}_{\omega}, n^{\rm in, v}_{\omega}, n^{\rm in, u}_{-\omega}$ since the 3  correlation terms initially vanish. Physically, this is a very legitimate assumption, as it means that the 3 modes are not correlated prior being scattered. In this case, using \eq{Bogcoef}, Eq.~\eqref{eq:noutcoutdef} gives
\begin{equation}
\label{eq:noutcoutSmatrix}
\begin{split}
n^{u}_{\omega}  = & \abs{\alpha_\omega}^2 n_{\omega}^{u, \rm in}+\abs{\beta_{\omega}}^2 (n_{-\omega}^{u, \rm in}+1)+\abs{ A_\omega}^2 n_{\omega}^{v, \rm in}, \\
n^{v}_{\omega} = & \abs{ \alpha^v_\omega}^2 n_{\omega}^{v, \rm in}+\abs{\tilde B_\omega}^2 (n_{-\omega}^{u, \rm in}+1)+\abs{\tilde A_\omega}^2 n_{\omega}^{u, \rm in} , \\
n^{u}_{-\omega} = & \abs{\alpha_{-\omega}}^2 n_{-\omega}^{u, \rm in} + \abs{\beta_{-\omega}}^2 (n_{\omega}^{u, \rm in}+1) \\
&+\abs{ B_\omega}^2 (n_{\omega}^{v, \rm in}+1) , \\
c^{uu}_\omega =  & \alpha_\omega \beta_{-\omega}^* (n_{\omega}^{u, \rm in}+\frac{1}{2})+\alpha_{-\omega} \beta_{\omega}^* (n_{-\omega}^{u, \rm in}+\frac{1}{2})\\
&+ A_\omega  B_\omega^* (n_{\omega}^{v, \rm in}+\frac{1}{2}) , \\
c^{uv}_\omega = & \tilde A_\omega \beta_{-\omega}^* (n_{\omega}^{u, \rm in}+\frac{1}{2})+\alpha_{-\omega} \tilde B_\omega^* (n_{-\omega}^{u, \rm in}+\frac{1}{2})\\
&+ \alpha^v_\omega  B_\omega^* (n_{\omega}^{v, \rm in}+\frac{1}{2}) , \\
d^{uv}_\omega = & \tilde A_\omega \alpha_{\omega}^* n_{\omega}^{u, \rm in}+\beta_{\omega} \tilde B_\omega^* (n_{-\omega}^{u, \rm in}+1)+ \alpha^v_\omega  A_\omega^* n_{\omega}^{v, \rm in} . 
\end{split}
\end{equation}
When working in the $in$ vacuum, $n^{u}_{\omega} = \abs{\beta_{\omega}}^2,\, n^{v}_{\omega} = \abs{\tilde B_{\omega}}^2 $ and $n^{u}_{-\omega} = \abs{\beta_{-\omega}}^2+ \abs{B_{\omega}}^2$ respectively give the mean number of the $u$ quanta {\it spontaneously} emitted to the right (the Hawking quanta), that of the $v$ quanta emitted to the left, and that of their negative energy partners. When the initial state is not the vacuum, the terms weighted by $n^{\rm in, a}_{\omega}$ give the {\it induced} contributions. One then sees that the norms $\abs{ A_\omega}^2, \abs{\tilde A_\omega}^2$ respectively quantify the grey-body factors, i.e. the reflexion of $v$ quanta into $u$ ones, and vice versa. 

Notice that unlike what is found for $2 \times 2$ $S$-matrices, one has  $\abs{\beta_{\omega}}^2 \neq \abs{\beta_{-\omega}}^2$, $\abs{B_\omega}^2 \neq \abs{\tilde B_\omega}^2$, and  $\abs{A_\omega}^2 \neq \abs{\tilde A_\omega}^2$. Yet, as shown by numerical simulations~\cite{Macher:2009nz,Macher:2009tw}, the relative difference between $\abs{\beta_{\omega}}^2$ and $\abs{\beta_{-\omega}}^2$ is generally small. Instead the differences $A_\omega - \tilde A_\omega$ and $B_\omega - \tilde B_\omega$ diverge in general when $\omega \to 0$. We shall return to this important point below. Notice finally that the coefficients  $\abs{\beta_{-\omega}}^2$ and $\abs{B_{\omega}}^2$ considered separately give the mean number of $u$ and $v$ quanta emitted by the corresponding white hole flow, see App.~\ref{app:whitehole} for more details. 

\subsection{Parameterization of the scattering}

We now show that, for stationary incoherent states, only four independent parameters of $S$ enter \eq{eq:NSepuv}. We shall work with the four squared norms $\abs{\beta_{\omega}}^2$, $\abs{\beta_{-\omega}}^2$, $\abs{ A_\omega}^2$ and $\abs{ B_\omega}^2$.

The $U(1,2)$ character of the $S$ matrix imposes the following relations from the normality of lines and columns
\begin{equation}
\label{eq:5modulusfixed}
\begin{split}
\abs{ \alpha^v_\omega}^2 &=  1+\abs{B_{\omega}}^2  - \abs{A_{\omega}}^2 , \\
\abs{\alpha_\omega}^2 &= 1+\abs{ \beta_\omega}^2 - \abs{A_\omega}^2 , \\
\abs{\alpha_{-\omega}}^2 &= 1+\abs{ \beta_{-\omega}}^2  + \abs{B_\omega}^2,  \\
\abs{\tilde A_\omega}^2 &=\abs{\beta_{-\omega}}^2  -\abs{ \beta_\omega}^2  + \abs{A_\omega}^2 , \\
\abs{\tilde B_\omega}^2 &=  \abs{ \beta_{-\omega}}^2   - \abs{\beta_{\omega}}^2 + \abs{B_\omega}^2,
\end{split}
\end{equation}
and from the orthogonality of the lines
\begin{subequations}
\label{eq:2linesortho}
\begin{align}
\label{eq:lign1ortho}
\tilde A_\omega \beta_{-\omega}^*-\alpha_{-\omega} \tilde B_\omega^* + \alpha^v_\omega  B_\omega^* =0,\\
\label{eq:lign2ortho}
\alpha_\omega \beta_{-\omega}^* -\alpha_{-\omega} \beta_{\omega}^* + A_\omega  B_\omega^* =0.
\end{align}
\end{subequations}
The number of real independent quantities is then reduces from $18$ ($9$ complex numbers) to $9$, which can be taken to be five phases and the above four norms. This choice is convenient because the $5$ phases drop out from \eq{eq:NSepuv}. 

In addition, Eqs.~\eqref{eq:2linesortho} and the positivity of the r.h.s. of Eqs.~\eqref{eq:5modulusfixed}, imply some inequality amongst the four norms.\footnote{
The origin of this fact is the following: Eqs.~\eqref{eq:2linesortho} define $2$ triangles in complex plane. Hence, one length cannot be larger than the sum of the two others. In addition to the $9$ real parameters of the $S$ matrix, one finds that there is an extra multiplicity $2$. It produces the symmetrical triangles with respect to the real axis. This extra multiplicity has no influence in the sequel since the initial state is incoherent.} 
These constraints are equivalent to 
\begin{subequations}
\begin{align}\label{eq:limitonABdomain}
\abs{A_{\omega}}^2 \leq 1+ \abs{B_{\omega}}^2 , \\
\label{eq:boundsonbeta}
\beta_{\omega}^{\rm min} \leq \abs{\beta_{\omega}} \leq \beta_{\omega}^{\rm max} , 
\end{align}
\end{subequations}
where
\begin{equation}
\begin{split}
\beta_{\omega}^{\rm min/max}&\doteq \abs{\frac{\abs{A_\omega B_\omega \alpha_{-\omega}} \pm\abs{\alpha_\omega^v\beta_{-\omega}}}{1+\abs{B_\omega}^2 }}.
\end{split}
\end{equation}
In this expression, $\alpha_{-\omega}$ and $\alpha_\omega^v$ are implicit expressions of $A_\omega, B_\omega$ and $\beta_{-\omega}$. To implement the right condition of Eq.~\eqref{eq:boundsonbeta} and reduce the number of independent norms to $3$, we impose
\begin{equation}
\label{eq:betaomimposed}
\begin{split}
\abs{\frac{\beta_{\omega}}{\beta_{-\omega}}} =\frac{\abs{\alpha_\omega^v} + \abs{A_\omega B_\omega }}{1+\abs{B_\omega}^2 } = \frac{1-\abs{A_\omega}^2 }{\abs{\alpha_\omega^v} - \abs{A_\omega B_\omega }}.
\end{split}
\end{equation}
The left condition of Eq.~\eqref{eq:boundsonbeta} is then equivalent to $ \abs{\beta_{\omega}\beta_{-\omega}}  \geq \abs{A_\omega B_\omega }^2 /4\abs{\alpha_\omega^v} $. To implement this inequality, we introduce $\abs{\beta^0_{\omega}}^2$ by
\begin{equation}
\label{eq:betaimposed2}
\begin{split}
 \abs{\beta_{\omega}\beta_{-\omega}} = \abs{\beta^0_{\omega}}^2 + \abs{A_\omega B_\omega }^2 /4\abs{\alpha_\omega^v}.
\end{split}
\end{equation}
When $A_\omega$ and $ B_\omega$ vanish, one has $\abs{\beta_{\omega}}^2 = \abs{\beta_{-\omega}}^2 =  \abs{\beta^0_{\omega}}^2$.

In conclusion, our parametrization of the relevant coefficients of the $S$ matrix is based on $\abs{ A_\omega}^2 ,\abs{ B_\omega}^2$ and $\abs{\beta^0_{\omega}}^2$. Eqs.~\eqref{eq:betaomimposed} and~\eqref{eq:betaimposed2} then fix $\abs{\beta_{\pm \omega}}^2$.

\subsection{Parametrization of dispersive spectra}

So far we worked at fixed $\omega$. To characterize the spectrum, 
we need to parametrize the $\omega$-dependence
of $\abs{ A_\omega}^2 ,\abs{ B_\omega}^2$ and $\abs{\beta^0_{\omega}}^2$. To this end, we consider the flow profile 
\begin{equation}
\label{eq:vpluscBH}
\begin{split}
v(x)  = - 1 + D \tanh \left ( \frac{\kappa x}{D} \right ).
\end{split}
\end{equation}
The parameter $D$ fixes the asymptotic values of $v+1$ on either side. In the present case they are equal and opposite. For more general asymmetric profiles we refer to~\cite{Finazzi:2012iu,Robertson:2011xp}. The frequency $\kappa$ fixes the surface gravity, and determines the temperature of the black hole radiation $T_H \doteq \kappa/(2\pi)$ in the {\it Hawking regime}, i.e. when dispersion effects are negligible because $\Lambda /\kappa \gg 1$. We work in units where $c = \hbar = k_B= 1$. When leaving this regime, numerical and analytical studies~\cite{Coutant:2011in} have established that $D$ also matters. In particular, when the coupling to the counter-propagating mode is small, i.e. if $\abs{ A_\omega}^2,\abs{ B_\omega}^2 \ll 1$, the spectrum of $u$ quanta spontaneously emitted $n^u_\omega = \abs{\beta^0_{\omega}}^2$ remains remarkably Planckian, even though the effective temperature, here after called $T_{\rm hor}$, is significantly modified. It is well approximated by~\cite{Robertson:2012ku,Finazzi:2012iu}
\begin{equation}
\begin{split}
T_{\rm hor} &\doteq T_H\tanh \left ( { T_\infty}/{T_H} \right ), \\ 
T_\infty &\doteq \frac{\Lambda D^{3/2}}{(2+D) \sqrt{2-D}}.
\label{Thor}
\end{split} 
\end{equation}
The effective temperature $T_{\rm hor}$ thus interpolates between the {\it Hawking regime} for low $T_H/T_\infty$, to the {\it dispersive regime} where it asymptotes to $T_\infty$ for $T_H/T_\infty \gg 1$. 

Numerical studies have also shown that  $\abs{ A_\omega}^2$ and $\abs{ B_\omega}^2$ are not fully determined by $\kappa, \Lambda, D$. They depend on the exact properties of the wave equation, and on the background profiles. Yet, they are generally smaller than $\abs{\beta^0_{\omega}}^2$, and remain finite for $\omega \to 0$. To implement these numerical observations, we shall work with 
\begin{equation}
\label{eq:ABbetaomegadep}
\begin{split}
\abs{ A_\omega}^2 &= \frac{2A^2}{\ep{\omega/T_{\rm hor}}+1},\quad
\abs{ B_\omega}^2 = \frac{2B^2}{\ep{\omega/T_{\rm hor}}+1},\\
\abs{\beta^0_{\omega}}^2&=\frac{1}{\ep{\omega/T_{\rm hor}}-1},
\end{split}
\end{equation}
where the constants $A^2$ and $B^2$ fix the overall norm of the two coupling between the counter-propagating mode with the Hawking mode and its partner.

It should be noticed that Eq.~\eqref{eq:ABbetaomegadep} and \eq{eq:5modulusfixed} correctly imply that only $A_\omega, B_\omega$ and $\alpha_\omega^v$ are regular in the limit $\omega \to 0$, whereas the squared norms of the 6 other coefficients diverge as $1/\omega$. It can be shown that this interesting property follows from the normalization in $1/\sqrt{\omega}$ of the low momentum modes.~\footnote{
We are grateful to Florent Michel for this explanation.} 

\subsection{Initial state}

To compute Eq.~\eqref{eq:noutcoutSmatrix} we also need the initial mean occupation numbers $n_\omega^{\rm in, a}$, where the superscript $a$ labels the three modes. As in Ref.~\cite{Macher:2009nz}, we assume that far from the horizon the initial state is a thermal bath at some global temperature $T_{\rm in}$ {\it in the frame of the fluid}. This means that the three $n_\omega^{\rm in, a}$ are given by
\begin{equation}
\label{eq:ninofOmega}
\begin{split}
n_\omega^{\rm in, a} = \frac{1}{\exp{(\Omega^{\rm in, a}_\omega/T_{\rm in})} -1},
\end{split}
\end{equation}
where $\Omega^{\rm in, a}_\omega$ is the asymptotic value of the co-moving frequency of the corresponding asymptotic $in$ mode. Its value is given by 
\begin{equation}
\begin{split}
\Omega^{\rm in, a}_\omega= \omega - v_{\rm as}^a k^{\rm in, a}_\omega ,
\end{split}
\end{equation}
where $k^{\rm in, a}_\omega $ is the corresponding wave vector, and where $v_{\rm as}^a$ is the asymptotic value of $v$ evaluated on the left or right side. In the present case, one has $v_{\rm as}^a = -1 \pm D $. The $-$ sign is associated to the $u$ modes, and the $+$ sign to the $v$ mode, see Fig.~\ref{fig:caracsuper}. 

In the low frequency limit, $\omega/\Lambda \ll 1 $, the expressions of $\Omega^{\rm in, a}_\omega$ can be analytically computed~\cite{Macher:2009nz}. Using them, one obtains
\begin{equation}
\label{eq:ninofmuT}
\begin{split}
n_{ \pm \omega}^{u,\rm in} &\sim \frac{1}{\exp\left [ (\mu \mp \omega) /T_{\rm in}^u  \right ] -1}, \\
n_{\omega}^{v, \rm in} &\sim \frac{1}{\exp\left ( \omega/T_{\rm in}^v  \right ) -1}.
\end{split}
\end{equation}
The chemical potential of $u$-modes, $\mu$, and the redshifted temperatures are 
\begin{equation}
\begin{split}
\frac{\mu}{ \Lambda } &= \frac{(1+ D) (D (2 + D))^{3/2}}{1+4 D +2 D^2},\\
{T_{\rm in}^u } &= {  T_{\rm in}}  \frac{ D(2 + D)}{1+4  D +2  D^2}, \\ 
T_{\rm in}^v &= T_{\rm in} ( 2  - D).
\label{Tuvin}
\end{split}
\end{equation}
The leading quantity governing $n_{ \pm \omega}^{u,\rm in}$ is $\mu /T_{\rm in}^u$. It scales as $\Lambda D^{1/2}/T_{\rm in}$. When $T_{\rm in} \ll \Lambda D^{1/2}$, the redshift is so important that the $u$-modes are effectively in their ground state, as in relativistic settings. 

\subsection{Summary} 
\label{sec:sixparameters}

Our parametrization of the final state, see \eq{eq:noutcoutSmatrix}, is based on $6$ dimensionless quantities, namely 
\begin{equation}
\label{eq:sixparameters}
\begin{split}
\omega/ T_{\rm hor},\  D,\ \Lambda/ \kappa,\ A,\ B, \mbox{ and } T_{\rm in } / T_{\rm hor}. 
\end{split}
\end{equation}
The first ratio is the frequency in the units of the effective temperature, which itself depends on the surface gravity $\kappa$, the dispersive wave-number $\Lambda$, and the hight of the velocity profile $D$, see \eq{Thor}. The parameters $A$ and $B$ respectively quantify the grey body factors and the pair creation of $uv$ pairs. The last ratio gives the initial temperature in the units of the effective temperature. 

Notice that these parameters are not independent, as $T_{\rm hor}$ depends on $D$. We have adopted this set, precisely because the residual dependence on $D$ at fixed $T_{\rm hor}$ is very weak. Hence, $D$ can be effectively fixed. As we shall see below, the other five parameters are all relevant. We believe they effectively provide a complete description of the system, at least when the flow profile is smooth enough. When it is not, the resonant effects~\cite{Zapata:2011ze,deNova:2012hm}, which are related to the black hole laser effect~\cite{Coutant:2009cu,Finazzi:2010nc,Finazzi:2010yq}, must be separately described.

\begin{figure*}
\begin{minipage}[t]{0.45\linewidth}
\includegraphics[width= 1 \linewidth]{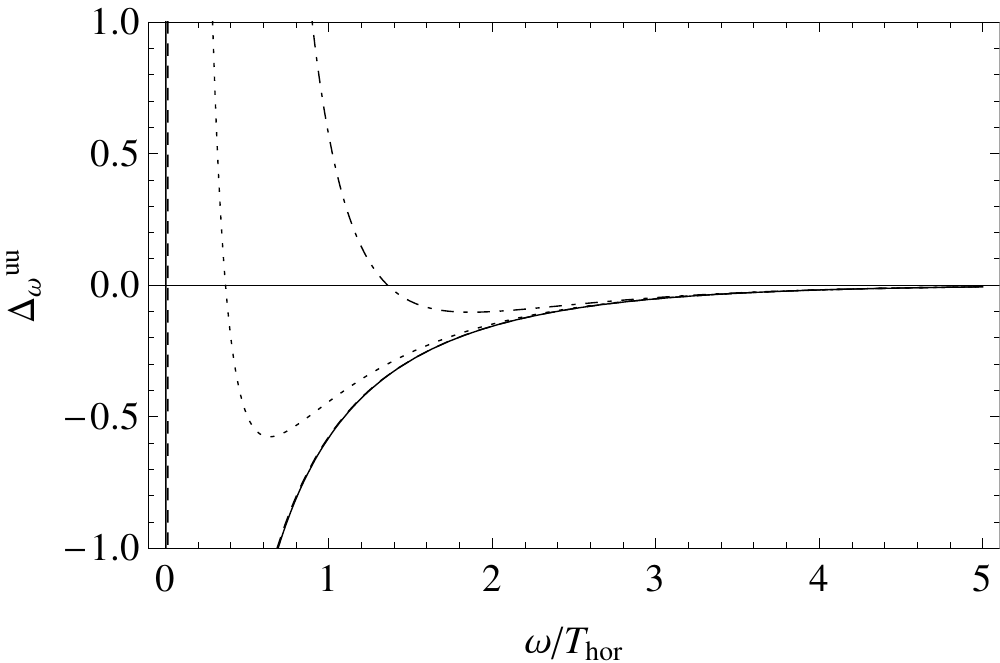}
\end{minipage}
\hspace{0.05\linewidth}
\begin{minipage}[t]{0.45\linewidth}
\includegraphics[width= 1 \linewidth]{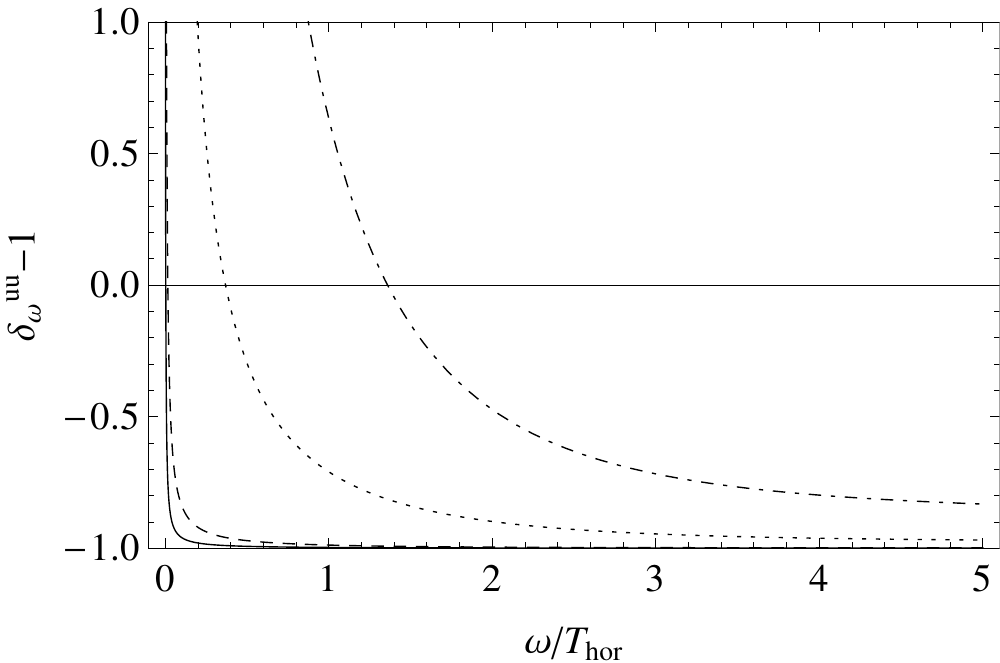}
\end{minipage}
\caption{The quantities $\Delta_\omega^{uu}$ of \eq{eq:Sepsimplified} (left panel) and $\delta_\omega^{uu}$ of \eq{eq:Nsepuvrelat} (right panel) are represented as functions of $\omega/T_{\rm hor}$ for a low initial temperature $T_{\rm in} = \Lambda/20$, for $\Lambda = 10 \kappa$, $D=1/2$, and for three values of $A = 4 B$, namely $B= 0.01$ (solid), $0.02$ (dash), $0.1$ (dot), and $0.25$ (dot-dash). One clearly sees that low frequency modes are separable, and that increasing $A = 4 B$ monotonously reduces the domain of non-separability.}
\label{fig:deltauofomegalowtemp}
\end{figure*}

\begin{figure*}
\begin{minipage}[t]{0.45\linewidth}
\includegraphics[width= 1 \linewidth]{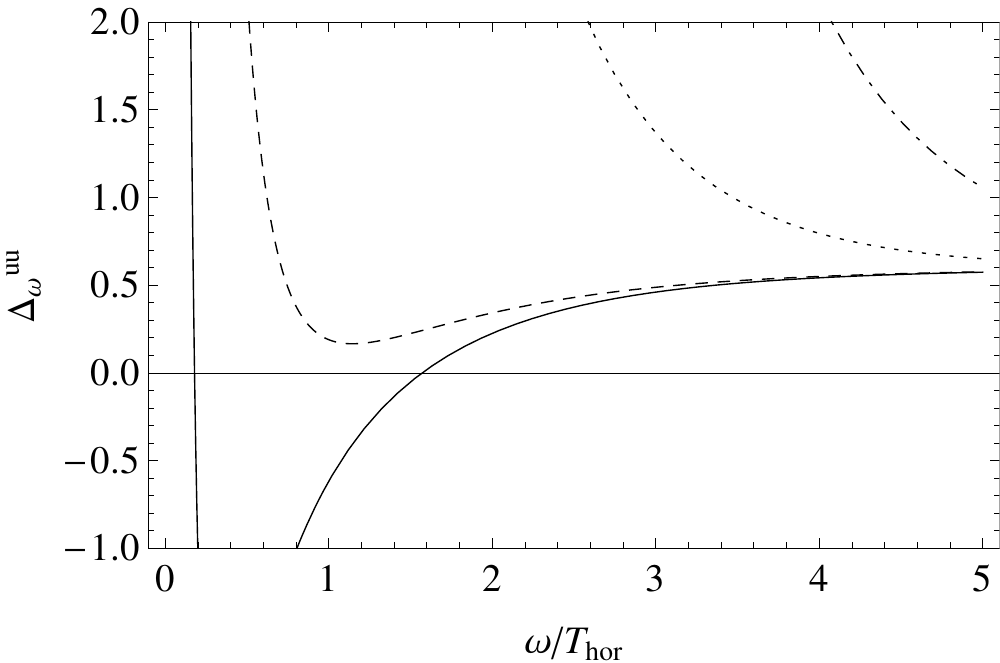}
\end{minipage}
\hspace{0.05\linewidth}
\begin{minipage}[t]{0.45\linewidth}
\includegraphics[width= 1 \linewidth]{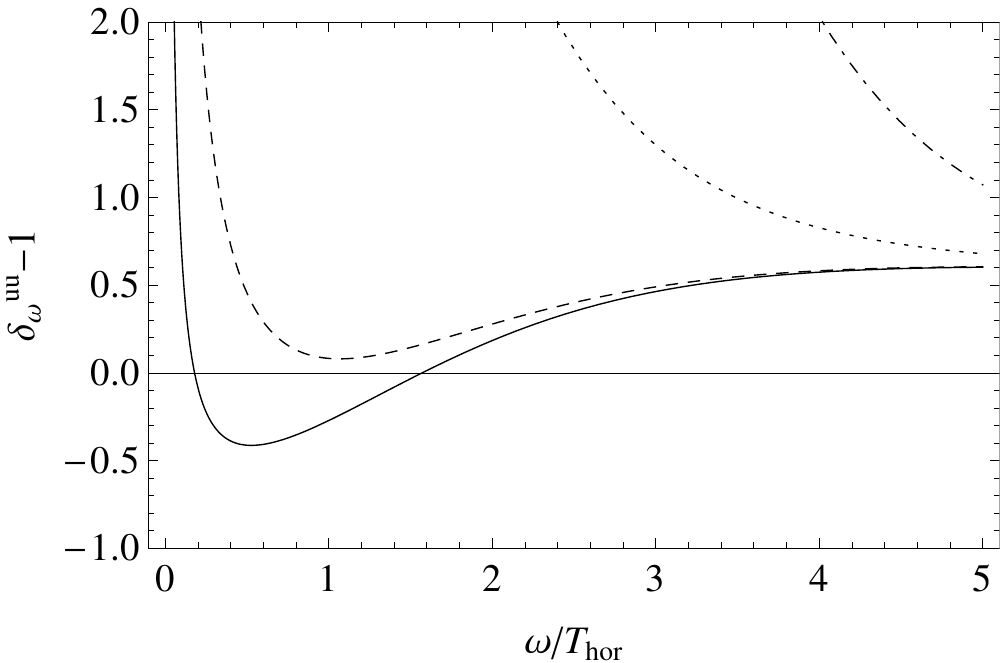}
\end{minipage}
\caption{We represent the same functions as in Fig.~\ref{fig:deltauofomegalowtemp}, for the same parameters, but for a higher initial temperature $T_{\rm in} = 2 \Lambda$. As expected, when compared to Fig.~\ref{fig:deltauofomegalowtemp}, one observes a reduction of the non-separability domains. As found at low temperature, increasing $A = 4 B$ still reduces the domain of non-separability. However, the high frequency sectors are now separable because $T_{\rm in}^u > 2 T_{\rm hor}$, as discussed below \eq{eq:Deltaulowomega2}. }
\label{fig:deltauofomegalargetemp}
\end{figure*}

\section{Domains of non-separability} 
\label{sec:domainsofnonsep}

Using Eqs.~\eqref{eq:5modulusfixed} and~\eqref{eq:2linesortho}, Eq.~\eqref{eq:NSepuv} can be expressed in terms of the  3 initial occupation numbers. In agreement with Eq.~(9) in Ref~\cite{deNova:2012hm}, we obtain
\begin{equation}
\label{eq:Sepsimplified}
\begin{split}
\Delta_\omega^{uu} =& \abs{\alpha_\omega^v}^2 n_{\omega}^{u, \rm in} n_{-\omega}^{u, \rm in} + \abs{\tilde B_\omega}^2 n_{\omega}^{u, \rm in} n_{\omega}^{v, \rm in} \\
&+ \abs{\tilde A_\omega}^2 n_{-\omega}^{u, \rm in} n_{\omega}^{v, \rm in} + \abs{B_\omega}^2 n_{\omega}^{u, \rm in} +  \abs{\beta_{-\omega}}^2 n_{\omega}^{v, \rm in} \\
&- \abs{\beta_{\omega}}^2 (1+ n_{\omega}^{v, \rm in} + n_{\omega}^{u, \rm in} + n_{-\omega}^{u, \rm in}) , \\
\Delta_\omega^{uv} =& \abs{A_\omega}^2 n_{\omega}^{u, \rm in} n_{-\omega}^{u, \rm in} + \abs{\beta_{\omega}}^2 n_{\omega}^{u, \rm in} n_{\omega}^{v, \rm in} \\
&+ \abs{\alpha_\omega}^2 n_{-\omega}^{u, \rm in} n_{\omega}^{v, \rm in} + \abs{B_\omega}^2 n_{\omega}^{u, \rm in} +  \abs{\beta_{-\omega}}^2 n_{\omega}^{v, \rm in} \\
&- \abs{\tilde B_\omega}^2 (1+ n_{\omega}^{v, \rm in} + n_{\omega}^{u, \rm in} + n_{-\omega}^{u, \rm in}) .
\end{split}
\end{equation}
We first notice that spontaneous pair production, i.e. $n_{\omega}^{a, \rm in} \equiv 0$, automatically leads to negative values for $\Delta_\omega^{uu} $ and $ \Delta_\omega^{uu} $, i.e. to non-separable states, see App.~\ref{app:separability}. We also notice that the above expressions are much more complicated than the corresponding ones in homogeneous and isotropic situations. In that case, because $n_\bk = n_{-\bk}$, one can work with a linear expression $n_\bk - |c_k|$ in the place of \eq{eq:NSepuv} which are quadratic in $n_{\rm in}^{\omega}$. As a result, one here looses the neat separation of the contributions of the spontaneous and the induced channels, see Eq.~(36) in \cite{Busch:2013sma}.

We finally notice that the maximum value of $\Delta_\omega^{uu}$ and $\Delta_\omega^{vv}$ is bounded by $ n_{-\omega}^{u}$. Indeed Heisenberg uncertainties guarantee [see B.4 in~\cite{Adamek:2013vw}]
\begin{equation}
\begin{split}
\abs{c^{u u/v}_\omega}^2 \leq n^u_{-\omega} (n^{u/v}_\omega  +1),
\end{split}
\end{equation}
for both the $uu$ and the $uv$ channels. It is thus useful to introduce the relative quantities~\cite{Campo:2008ju}
\begin{equation}
\label{eq:Nsepuvrelat}
\begin{split}
\delta_\omega^{uu} \doteq  \frac{\Delta_\omega^{uu}}{ n_{-\omega}^{u} }+1,  \quad
\delta_\omega^{uv} \doteq  \frac{\Delta_\omega^{uv}}{  n_{-\omega}^{u} }+1 .
\end{split}
\end{equation}
which are both positive, irrespectively of the state $\hat \rho$.

In what follows, we study the domains of negativity of $\Delta_\omega^{uu}$ and $\Delta_\omega^{vv}$ by making use of the parametrization of Sec.~\ref{sec:system}. In the body of the text, we consider the dispersion relation of Eq.~\eqref{eq:disprelsuper}. The sub-luminal case is briefly studied in Appendix~\ref{app:sublum}. To identify the domains of non-separability, we shall mainly use figures.

\subsection{Non-separability of \texorpdfstring{\boldmath{$uu$}}{uu} pairs}

\subsubsection{The dependence in \texorpdfstring{$\omega$}{omega}}

In Fig.~\ref{fig:deltauofomegalowtemp}, we study the $\omega/T_{\rm hor}$ dependence of $\Delta_\omega^{uu}$ and $\delta_\omega^{uu}$ in the low initial temperature regime, for $T_{\rm in} = \Lambda/20$. We consider three  different values of the coefficients $A,B$ of Eq.~\eqref{eq:ABbetaomegadep}, with a fixed ratio $A/B = 4$. At large frequency, $\omega/T_{\rm hor} \gtrsim 2$, the state is non separable independently of the values of $A$ and $B$. On the other hand, at low frequency, the state is always separable, even though decreasing $A$ and $B$ clearly increases the domain of non separability. The minimum value of $\Delta_\omega^{uu}$ is reached for $\omega/T_{\rm hor}\sim 1$. Instead, the minimum of $\delta_\omega^{uu}$ is reached for $\omega \to \infty$. In brief, for low initial temperatures, the low frequency sector contains many pairs but they are separable, the high frequency sector contains very few pairs which are highly non separable, and the intermediate regime contains few of them which are barely non separable. 

\begin{figure*}
\begin{minipage}[t]{0.45\linewidth}
\includegraphics[width= 1 \linewidth]{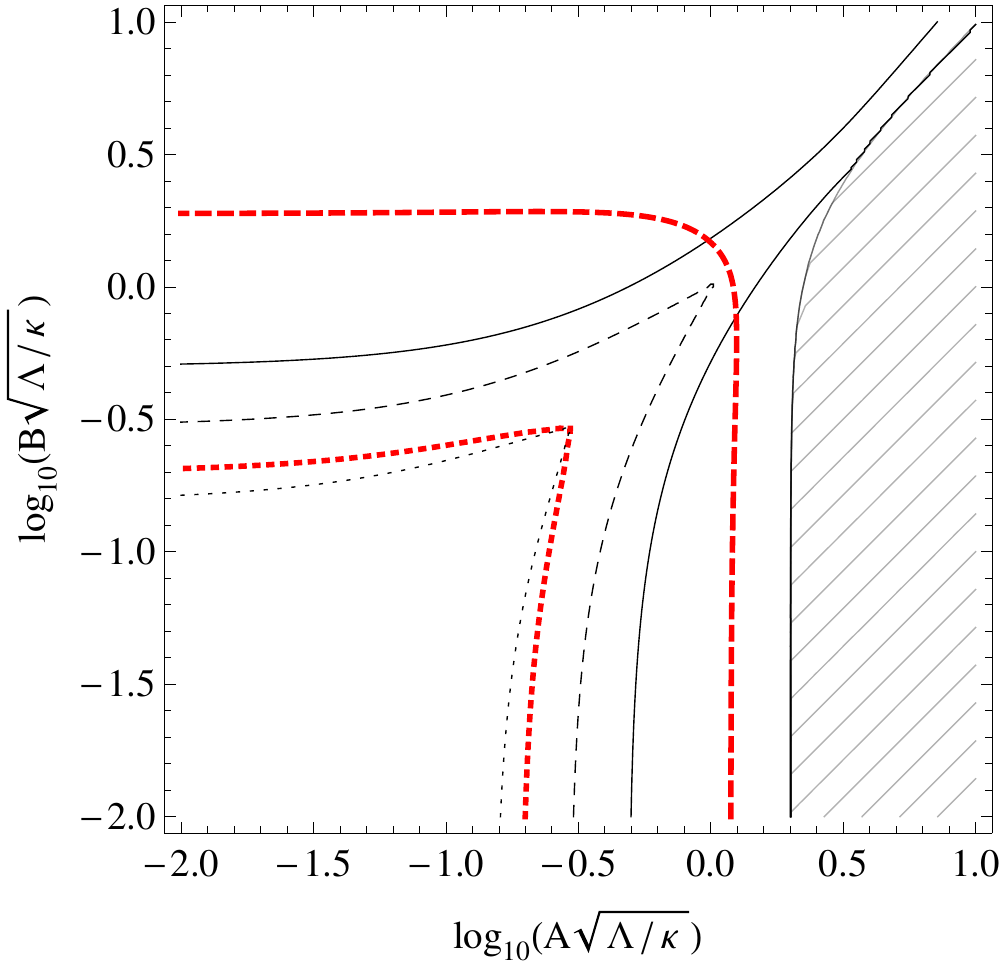}
\caption{The minimum value of $\Delta_\omega^{uu}$ over $\omega$ in the plane of $\log_{10} A \sqrt{\Lambda/\kappa}$, $\log_{10} B\sqrt{\Lambda/\kappa}$, for three initial temperatures $ 5 T_{\rm in}^u /2 \mu = 1/3$ (Solid), $1$ (Dashed) or $3$ (Dotted), and for $D = 1/2$, $\kappa /\Lambda = 1/4$. The thick red line is $\min \Delta_\omega^{uu} = 0$, and indicates the limit of non-separability. The black line gives $\min \Delta_\omega^{uu} = -0.5$. It indicates the domain where the non-separability is significant. We have $T_H / T_{\infty} \sim 1/3$, so we work at the edge of the {\it Hawking regime}. The dashed region represents the forbidden region where $\abs{\alpha^v}^2 = 1+\abs{B}^2-\abs{A}^2 <0$. }
\label{fig:minsepuhawk}
\end{minipage}
\hspace{0.05\linewidth}
\begin{minipage}[t]{0.45\linewidth}
\includegraphics[width=1\linewidth]{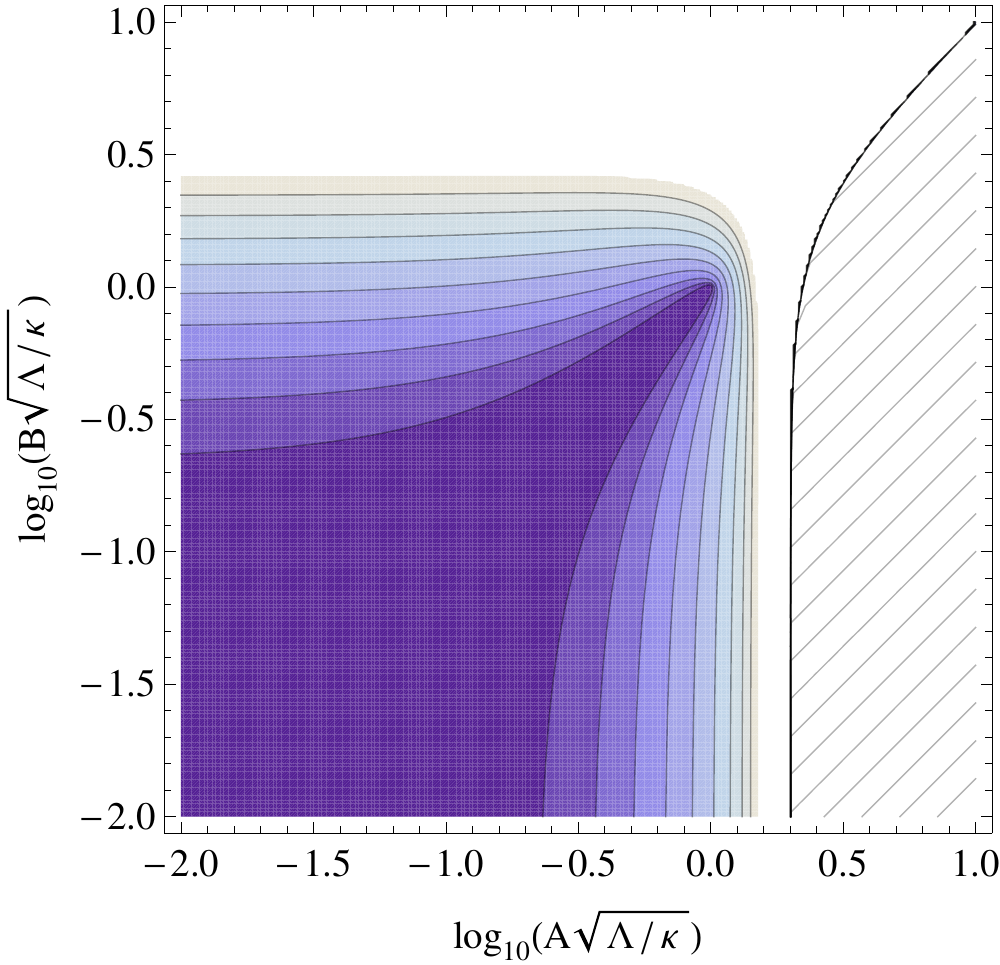}
\caption{The value of $\omega/T_{\rm hor}$ that minimizes $\Delta_\omega^{uu}$ in the same plane as in Fig.~\ref{fig:minsepuhawk}, for the same parameters, and for the intermediate temperature $5 T_{\rm in}^u /2 \mu = 1$. The lines of constant $\omega/T_{\rm hor}$ go by step of $1/2$ from $1/2$ (dark blue) to $4.5$ (clear blue). The most relevant sector is $\omega/T_{\rm hor}\lesssim 1$, see Eq.~\eqref{eq:ABbetaomegadep}. } 
\label{fig:omegaminu}
\end{minipage}
\end{figure*}

In Fig.~\ref{fig:deltauofomegalargetemp}, we study the same functions for a much larger initial temperature: $T_{\rm in} = 2 \Lambda$. In this case, the induced effects are much more important than above. As a result, the final state becomes separable even at large frequency. In fact, according to the value of $A$ and $B$, three different regimes show up. When $A,B$ are low enough, there still exists a finite range in $\omega$ where the state is non separable. When increasing $A,B$, this domain disappears, but $\Delta_\omega^{uu}$ still possesses a local minimum. When further increasing $A$ and $B$, $\Delta_\omega^{uu}$ becomes a monotonically decreasing function of $\omega$. Hence, as expected, increasing the initial temperature severely restricts the non-separability of the state, or even completely suppresses it. 

It is of value to study analytically the asymptotic behaviors. The infra-red behavior is dominated by the values of $A$ and $B$, as can be seen from 
\begin{equation}
\label{eq:Deltaulowomega}
\begin{split}
\Delta_\omega^{uu} \underset{\omega \to 0}{\sim } \frac{T_{\rm hor} T_{\rm in}^v}{\omega^2} \left ( n_{\omega}^{u, \rm in} + n_{-\omega}^{u, \rm in}+1 \right ) \left (  \gamma_+  B - A   \gamma_- \right )^2 ,
\end{split}
\end{equation}
where $\gamma_\pm$ is the limit of $\sqrt{\omega \abs{\beta_{\pm \omega}}^2 / T_{\rm hor} }$ for $\omega \to 0$. Eq.~\eqref{eq:Deltaulowomega} diverges as $\omega \to 0$ and is positive defined. This implies the separability of the low frequency regime.~\footnote{There is a noticeable exception: when $A=B$, the leading divergence in $1/\omega^2$ is absent. As a result, the domain of non-separability further extends at low frequency. This is the case studied in~\cite{Unruh:1994je,Brout:1995wp,Macher:2009tw} where high frequency dispersion is added on the 2-dimensional massless scalar equation. } 
In the large frequency regime $ \mu > \omega  \gg  T_{\rm in},T_{\rm hor} $, we obtain 
\begin{equation}
\begin{split}
\Delta_\omega^{uu} \sim \frac{e^{-2 \mu /T_{\rm in}^u}-  e^{-{\omega }/{T_{\rm hor}}}}{1-e^{-\abs{\mu-\omega  }/T_{\rm in}^u}}.
\label{eq:Deltaulowomega2}
\end{split}
\end{equation}
This is negative when $\omega T_{\rm in}^u \lesssim 2 \mu  T_{\rm hor}$. Hence, since $\omega<\mu$, the UV sector is non separable if $2  T_{\rm hor} \gtrsim T_{\rm in}^u $. With $D=1/2$ and $\Lambda = 10 \kappa$, this limit corresponds to $ T_{\rm in} \sim \Lambda/11$, independently of the values of $A$ and $B$. In addition, we observed that changing $D$ at fixed $T_{\rm hor}$ and $T^u_{in}$ of \eq{Tuvin} has basically no effect.

In brief, the state in the infra-red sector is generically separable because of the divergent contribution governed by the coefficients $A$ and $B$. Instead, the separability in the ultraviolet sector critically depends on $T_{\rm in}^u/T_{\rm hor}$. The intermediate regime is non separable if $A,B$ are low enough. 

Having characterized how $\Delta_\omega^{uu}$ depends on $\omega/T_{\rm hor}$, we now study how it depends on $A,B,T_{\rm hor}$ and $T_{\rm in}$. We shall establish that only two types of behaviors are found, depending on the ratio $T_H/T_{\infty}$, see \eq{Thor}. In this we extend what was found in the spectral analysis of Ref.~\cite{Finazzi:2012iu}. When $T_H/T_{\infty} \lesssim 1/3$, one lives in the Hawking regime, with small dispersive effects. Instead, when  $T_H/T_{\infty} \gtrsim 3$, one finds the dispersive regime where the surface gravity plays no significant role. 

\subsubsection{Hawking regime} 

\begin{figure}
\includegraphics[width= 1 \linewidth]{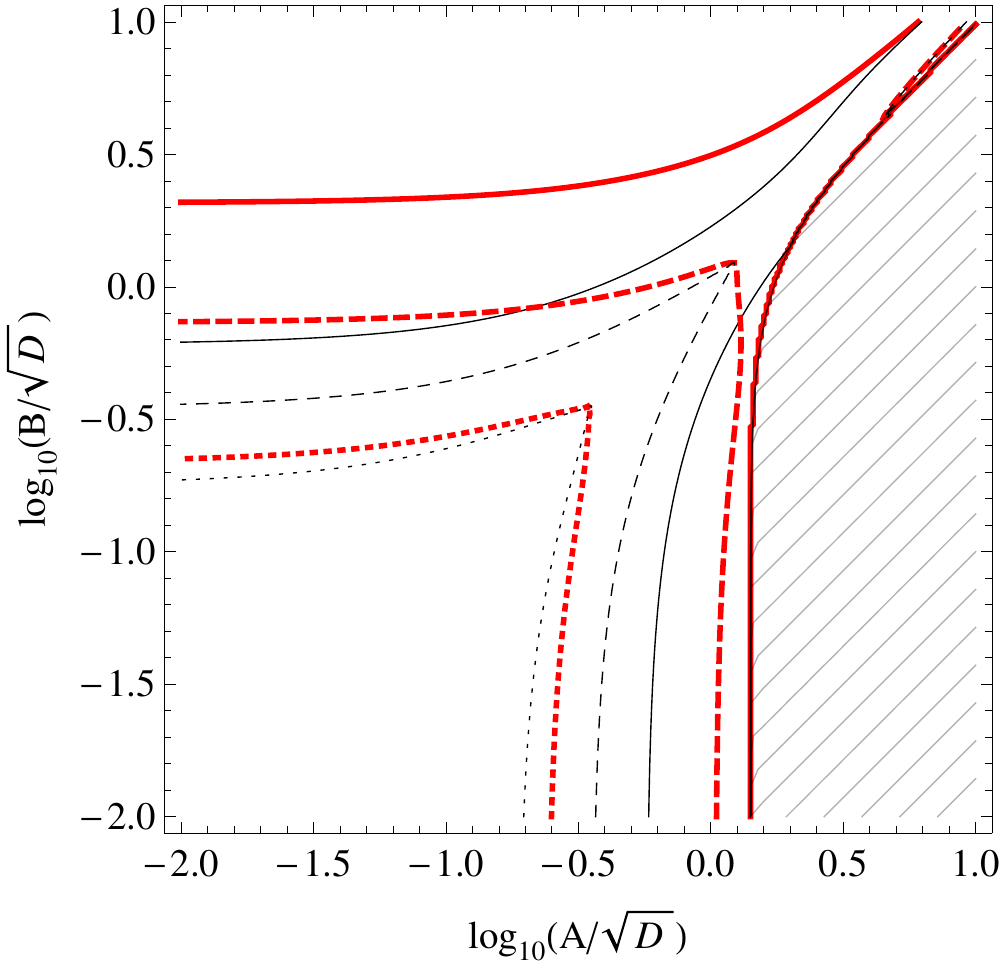}
\caption{As in Fig.~\ref{fig:minsepuhawk}, we represent the minimum value of $\Delta_\omega^{uu}$ for three initial temperatures $ 5 T_{\rm in}^u /2\mu = 1/3$ (Solid), $1$ (Dashed) or $3$ (Dotted), and for $D = 1/2$. Here, we work in the {\it dispersive regime} since $\kappa /\Lambda = 10/4 $, and $T_H / T_{\infty} \sim 3$. As explained in the text, the two coordinates now are $\log_{10} A/\sqrt{D}$ and $\log_{10} B/\sqrt{D}$. The black line represents $\min \Delta_\omega^{uu} = -0.5$, and the thick red line $\min \Delta_\omega^{uu} = 0$. Notice the similarity of the present figure with Fig.~\ref{fig:minsepuhawk}. It indicates that the cross-over, around $\Lambda/\kappa D \sim 1$, from the Hawking regime to the dispersive one is smooth.}
\label{fig:minsepudisp}
\end{figure}

We first work at the edge of the Hawking regime, with $T_H/T_{\infty} = 1/3$. Reducing this ratio, which means reducing $\kappa/\Lambda$, does not affect the properties of Fig.~\ref{fig:minsepuhawk}. Hence what follows applies to the entire Hawking regime.

To eliminate $\omega$, we consider the minimum value of $\Delta_\omega^{uu}$ for $\omega < 5 T_{\rm hor}$. (It is pointless to consider higher values since the pair production rates are exponentially suppressed in that regime.) We shall consider two values of the minimum, namely $\min_\omega \Delta_\omega^{uu}= 0$ and $ = -0.5$. The first one gives the limit of non-separability, whereas the second curve indicates the domain where the non-separability is significant, and therefore more likely to be observed in an experiment. In Fig.~\ref{fig:minsepuhawk}, both curves $\min_\omega \Delta_\omega^{uu}= 0$ and $ = -0.5$ are represented in the plane of $\log_{10} (A\sqrt{\Lambda/\kappa})$ and $\log_{10}( B\sqrt{\Lambda/\kappa})$, and for three different initial temperatures, namely $ 5 T_{\rm in}^u /2 \mu  = 1/3$, $1$ or $3$. After several tries, we have adopted these axis and this parametrization of the initial temperature, because changing $D$ and $\Lambda$ at fixed 
\begin{equation}
\begin{split}
\frac{ T_{\rm in}^u }{2\mu}, \, {A}{\sqrt{\Lambda/\kappa} }, \,  {B}{\sqrt{\Lambda/\kappa} },
\label{3hawk}
\end{split}
\end{equation}
has no significant influence on the curves. This means that in the Hawking regime, the minimal value of $\Delta_{uu}$ only depends on these three composite scales. This is our first important result. 

The other lesson from Fig.~\ref{fig:minsepuhawk} is that $A$ and $B$ should be both smaller than $\sim \sqrt{T_{\rm in}^u /{6\mu}} \times\sqrt{\kappa/\Lambda}$ for the state to be significantly entangled, i.e. $\Delta_\omega^{uu} < - .5$. When this condition is met, the state can be found entangled even when the initial temperature $T_{\rm in}$ is significantly larger than the horizon temperature $T_{\rm hor}$. To give an example, when $T_{\rm in} = \Lambda = 10 T_{\rm hor}$, the state is non-separable if $ A, B \lesssim 1/10 $. In addition, one also sees that $A\sim B$ enhances the non-separability of the state. This is because the $1/\omega^2$ divergence of \eq{eq:Deltaulowomega} is reduced when $A\sim B$. 

To complete the information and also guide future experiments, in Fig.~\ref{fig:omegaminu} we represent the value of $\omega/T_{\rm hor}$ that minimizes $\Delta_\omega^{uu}$ for the same parameters as those of Fig.~\ref{fig:minsepuhawk}, and for the middle temperature $T_{\rm in}^u / \mu = 2/5$. Notice that a rough characterization of the curves can be obtained by considering the asymptotic behaviors of Eqs.~\eqref{eq:Deltaulowomega} and~\eqref{eq:Deltaulowomega2}, and by minimizing their sum. The symmetry with respect to interchanging $A$ and $B$ is then explained. In addition, when increasing the initial temperature $T_{\rm in}$, we learn that one should increase the value of $\omega/T_{\rm hor}$ in order to minimize $\Delta_\omega^{uu}$. Roughly speaking, one gets $(\omega/T_{\rm hor})^3\sim T_{\rm in}/T_{\rm hor}$.

\begin{figure*}
\begin{minipage}[t]{0.45\linewidth}
\includegraphics[width= 1 \linewidth]{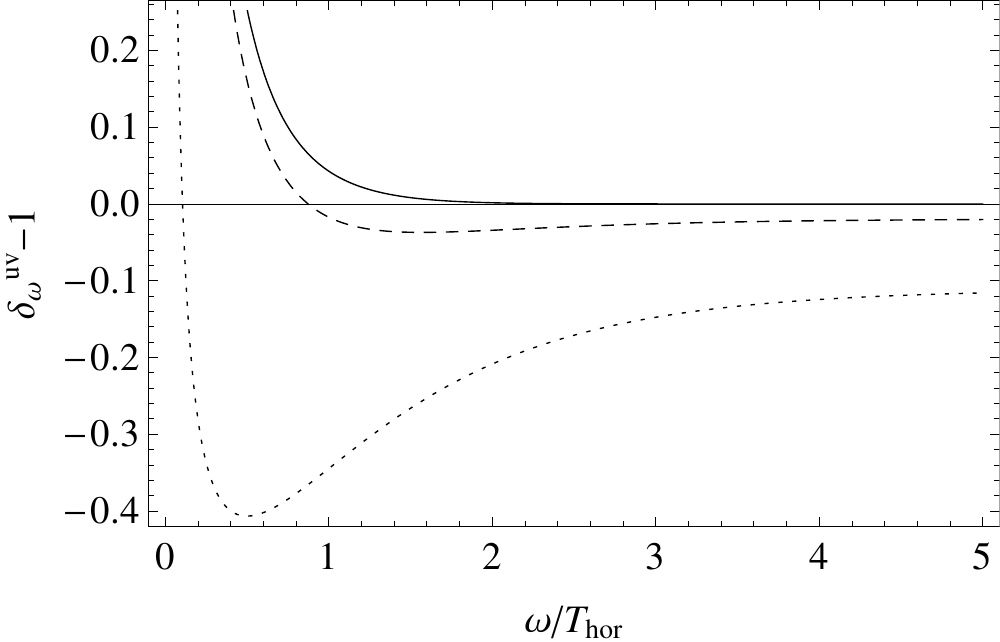}
\caption{The relative quantity $\delta_\omega^{uv}$ of Eq.~\eqref{eq:Nsepuvrelat} as a function of $\omega/T_{\rm hor}$ for a low temperature $T_{\rm in} = \Lambda /300 $, for three values of $B= 0.01$ (solid), $0.1$ (dash), and $0.25$ (dot), for $A = 4 B$, $\Lambda = 10 \kappa$, and $D=1/2$. For $uv$ pairs, increasing $B$ now increases the non-separability since $B^2$ governs their creation rate.}
\label{fig:vlowtempofomega}
\end{minipage}
\hspace{0.05\linewidth}
\begin{minipage}[t]{0.45\linewidth}
\includegraphics[width= 1 \linewidth]{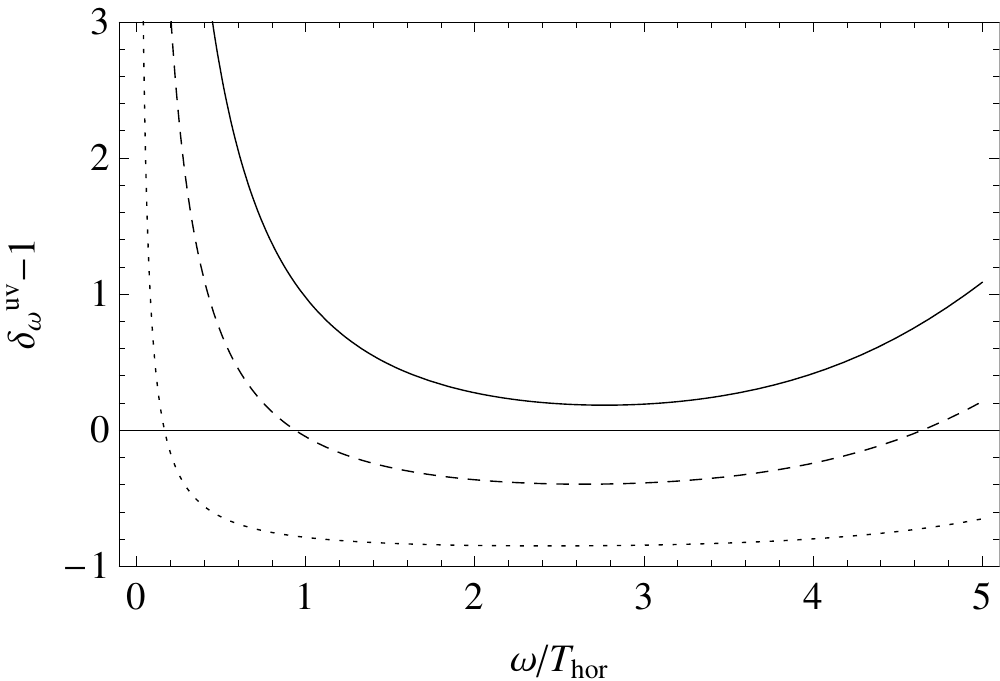}
\caption{The same relative quantity $\delta_\omega^{uv}$ as a function of $\omega/T_{\rm hor}$ for a high temperature $T_{\rm in} = \Lambda /3 $, for three values of $B= 0.1$ (solid), $1$ (dash), and $2$ (dot), for  $A = B/4$, $\Lambda = 10 \kappa$, and $D=1/2$. With respect to Fig~\ref{fig:vlowtempofomega}, the values of $B$ have been increased by a factor of $\sim 10$. }
\label{fig:vlargetempofomega}
\end{minipage}
\end{figure*}

\subsubsection{The dispersive regime}

We now proceed in the same way for the dispersive regime. We work at the edge of this domain with $T_H/T_{\infty} = 3$. We have verified that what follows applies for $T_H/T_{\infty} > 3$. 

As for the Hawking regime, we extract the $\omega$ dependence by taking the minimum of $\Delta_\omega^{uu}$ over $\omega$, for $\omega < \omega_{\max}$, where $\omega_{\max}$ is the maximum value for which the negative norm mode exists~\cite{Macher:2009tw}. Since we are in the dispersive regime, the approximate expression of Eq.~\eqref{eq:ninofmuT} is no longer valid, even though $\mu$ and $T_{\rm in}^u$ of Eq.~\eqref{Tuvin} are still well defined. We thus use the exact expression of Eq.~\eqref{eq:ninofOmega} in this Section. 

As in Fig.~\ref{fig:minsepuhawk}, in Fig.~\ref{fig:minsepudisp} we draw the constant values $\min_\omega \Delta_\omega^{uu}$ equal $0$ and $-0.5$, to respectively get the non-separability, and the significantly non-separable, domains. In the present case, the axes have been chosen to be $\log_{10} (A/\sqrt{D}), \, \log_{10} (B/\sqrt{D})$, because, when adopting them, we observed that changing $D$ and $\Lambda$ at fixed $T_{\rm in}^u /\mu$ and $\{A,B\}/\sqrt{D}$ has no significant effect. As a result, in the dispersive regime, the minimal value of $\Delta_{uu}$ only depends on the following three composite scales 
\begin{equation}
\begin{split}
\frac{T_{\rm in}^u}{2 \mu }, \, \frac{A}{\sqrt{D }}, \,  \frac{B}{\sqrt{D} }. 
\label{3disp}
\end{split}
\end{equation}
This is the second important result of this paper. Notice that these three ratios differ from those of \eq{3hawk}. 
We also notice that Fig.~\ref{fig:minsepudisp} is very similar to Fig.~\ref{fig:minsepuhawk}. This means that the cross-over, around $\Lambda/\kappa D \sim 1$, from the Hawking regime to the dispersive one is rather smooth. 

As a result, when taken together, Figs.~\ref{fig:minsepuhawk} and~\ref{fig:minsepudisp} offer a full characterization of the non-separability domains when the initial temperature belongs to the domain $0.1\lesssim T_{\rm in}/\mu \lesssim 1  $.\footnote{We remind the reader that the initial temperature in polariton system under coherent pumping is $T_{\rm in}= \Lambda $~\cite{Busch:2013gna} which belongs to this range.} The main conclusion of this Section is that separability of the state depends mainly on three quantities. The first one is the initial temperature in the unit of the chemical potential $\mu$. This was expected since this ratio governs the initial distribution of $u$-quasi-particles. The other two are the $A$ and $B$ parameters encoding the coupling with the third mode. In order to obtain domains with well defined scaling properties, these dimensionless parameters should be rescaled by $\sqrt{\Lambda/\kappa}$ in the Hawking regime, and by $\sqrt{D}$ in the dispersive one. 

\subsection{Non-separability of \texorpdfstring{\boldmath{$uv$}}{uv} pairs}

\subsubsection{The dependence in \texorpdfstring{$\omega$}{omega}}
 
Because the analysis is rather similar to that of the previous Section, we only give the main results. In Fig.~\ref{fig:vlowtempofomega}, we first study $\delta_\omega^{uv}$ as a function of $\omega/T_{\rm hor}$ for three different values of $A,B$. As was found Fig.~\ref{fig:deltauofomegalowtemp}, we observe that the low frequency sector is always separable. 

When increasing the initial temperature, as expected, the value of $\delta_\omega^{uv}$ increases and the state becomes separable for all $\omega$. We then need to increase the pair creation rate $B^2$ to get non separable states, see Fig.~\ref{fig:vlargetempofomega}. We observe that the low frequency sector remains as it was at low temperature. On the contrary, for high frequency, the state is now separable. Yet, when $B$ is large enough, there exists an intermediate regime where the state remains non separable. In this regime, the non separability depends on a competition between the coupling $B$ and the initial temperature.

These observations can be verified analytically. First, the low frequency behavior is
\begin{equation}
\label{eq:Deltauvlowom}
\begin{split}
\Delta_\omega^{uv} \underset{\omega \to 0}{\sim}  \frac{ T_{\rm hor} T_v }{\omega^2 } \gamma_+^2 \left ( n_{\omega}^{u, \rm in} + n_{-\omega}^{u, \rm in}+1 \right ).
\end{split}
\end{equation}
We obtain a behavior similar to that of $uu$ pairs given in \eq{eq:Deltaulowomega}. However, in the present case, $\Delta_\omega^{uv} $ remains positive even when $A=B$. Second, at large frequency, the behavior of $\Delta_\omega^{uv}$ is very different to that of $\Delta_\omega^{uu}$. Indeed, when $\mu> \omega \gg T_{\rm in},T_{\rm hor}$, we have
\begin{equation}
\label{eq:Deltauvlargeom}
\begin{split}
\Delta_\omega^{uv} \sim&  \frac{ e^{-\omega /T_{\rm in}^v }+\ep{ \abs{\omega -\mu }/T_{\rm in}^u-\omega /T_{\rm hor}} (e^{-\omega /T_{\rm in}^v } - 2 B^2) }{e^{ \abs{\omega -\mu }/T_{\rm in}^u}-1}.\\
\end{split}
\end{equation}
The source of non separability is the term proportional to $B^2$. This makes perfectly sense since $B^2$ fixes the production of $uv$ pairs. With more precision, the large frequency ($\omega = \mu$) behavior is non separable only if
\begin{equation}
\label{eq:vUVnonsep}
\begin{split}
2 B^2 \gtrsim \ep{\mu/T_{\rm hor} - \mu /T_{\rm in}^v},
\end{split}
\end{equation}
which requires a very low initial temperature in order to be satisfied. When Eq.~\eqref{eq:vUVnonsep} is not fulfilled, the state is separable at large $\omega$. However, non separability is possible when
\begin{equation}
\begin{split}
2 B^2 \gtrsim  \ep{ - \mu T_{\rm hor} / T_{\rm in}^v (T_{\rm hor}+ T_{\rm in}^u )}
\end{split}
\end{equation}
for frequencies obeying 
\begin{equation}
\begin{split}
\omega \lesssim \frac{ \mu + T_{\rm in}^u \log(2B^2) }{1 + T_{\rm in}^u /T_{\rm hor}- T_{\rm in}^u/T_{\rm in}^v} .
\end{split}
\end{equation}
This is the order of magnitude that we observe in fig.~\ref{fig:vlargetempofomega}.

\begin{figure*}
\begin{minipage}[t]{0.45\linewidth}
\includegraphics[width= 1 \linewidth]{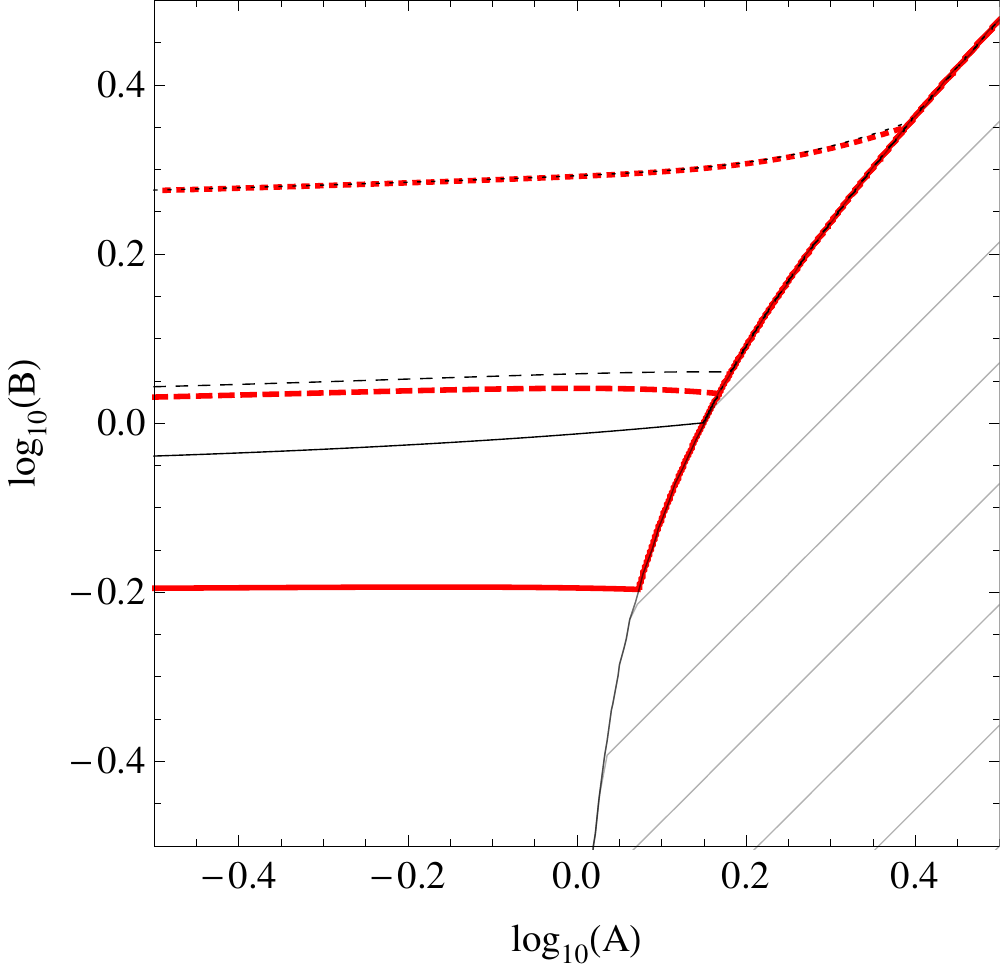}
\caption{The minimum over $\omega$ of $\Delta_\omega^{uv}$ for $ T = 1/3 \Lambda \sqrt{D}$ (Dotted), $\Lambda\sqrt{D}$(Dashed) or $3 \Lambda \sqrt{D}$ (solid), $\Lambda = 4 \kappa$, $D=1/2$. The dashed region represent the region with $\abs{\alpha^v}^2 = 1+\abs{B}^2-\abs{A}^2 <0$. The line $\min \Delta = 0$ is indicated in thick red. The line $\min \Delta = -0.5$ is indicated in black. }
\label{fig:minsepv}
\end{minipage}
\hspace{0.05\linewidth}
\begin{minipage}[t]{0.45\linewidth}
\includegraphics[width= 1 \linewidth]{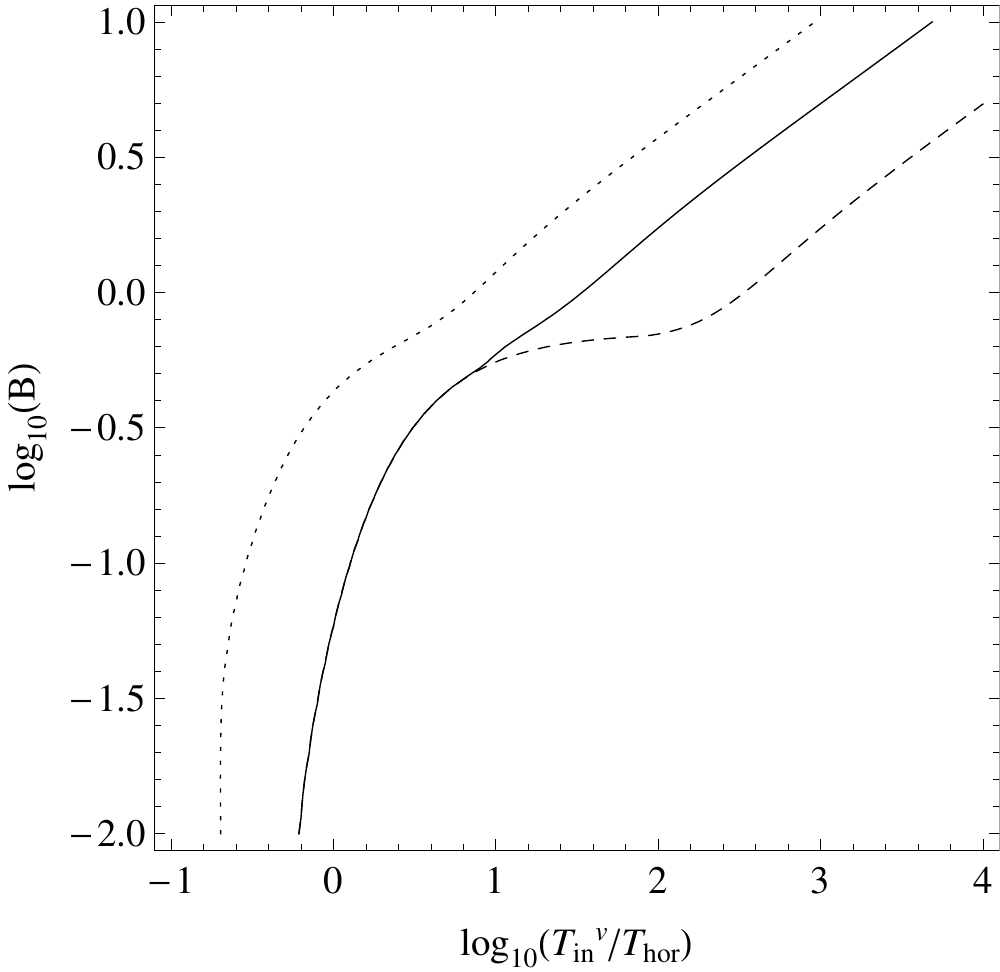}
\caption{The threshold of $uv$ non-separability $\Delta_\omega^{uv} = 0$ in the $T_{in}^v/T_{\rm hor},B$ plane for $3$ values of $\Lambda\sqrt{D}/\kappa$, i.e., $40$ (dashed), $4$  (solid), $0.4$ (dotted). In the Hawking regime, there exists a critical temperature, which controls the separability of the $UV$ sector. This critical temperature disappears in dispersive regime. At larger temperature, non-separability is found if $B^2> T_{\rm in}^v / \Lambda\sqrt{D}$. When going further in dispersive regime, $\Delta_\omega^{uv} = 0$ no longer evolves, and remains along dotted curve. } 
\label{fig:enrab}
\end{minipage}
\end{figure*}

\subsubsection{The parametric dependence}

As for $uu$ pairs, we now determine what is the domain of the parameter space where the state is non-separable. We represent in Fig.~\ref{fig:minsepv} the minimum over $\omega$ of $\Delta_\omega^{uv}$ for different temperatures in the $A,B$ plane. The first observation is that $A$ plays no role. This can be seen from the asymptotic behaviors of Eqs.~\eqref{eq:Deltauvlowom} and~\eqref{eq:Deltauvlargeom}. A closer analysis reveals that the relevant parameters governing the non-separability of $uv$ pairs are 
\begin{equation}
\begin{split}
\frac{T_{\rm in}^v}{T_{\rm hor}}, B, \frac{\Lambda \sqrt{D}}{\kappa}.
\end{split}
\end{equation}
This is the third important result of this paper.

Given that observation, we represent in fig.~\ref{fig:enrab} the non-separability threshold $\Delta_\omega^{uv} = 0$ in the $B$, $T_{\rm in}/T_{\rm hor}$ plane for different values of $\Lambda/ \kappa$. We observe first, that in the Hawking regime, there is a critical temperature $T_{\rm in}^{\rm crit} \sim  T_{\rm hor} $ below which the state is always non separable. This limit is due to the $UV$ behavior of the spectrum. Indeed, we see from Eq.~\eqref{eq:vUVnonsep} that when $\mu$ is large (i.e., deep in the Hawking regime) and $T_{\rm in}^v < T_{\rm hor}$, the states with $\omega \sim \mu$ are non separable for all value of $B$. The second observation is that this critical temperature decreases as we leave the Hawking regime. This is because the non separable regime $\omega \gg T_{\rm hor}$ no longer exists when $T_{\rm hor} \sim \omega_{\max}$. At higher temperature, the non separability criterion becomes $ B^2 \gtrsim T_{\rm in}/\Lambda$.

To summarize, the state is non separable when $T_{\rm in}^v \lesssim T_{\rm hor} $ or $T_{\rm in}^v \lesssim B^2 \Lambda $. 

\section{Conclusions}

We analyzed the strength of the correlations characterizing the two types of pairs that are emitted by a stationary black hole flow. To distinguish classical correlations associated with stimulated effects from genuine quantum entanglement due to spontaneous effects, we worked at fixed $\omega$, and used the criterion of non-separability of the state which implies that one of the differences of \eq{eq:NSepuv} should be negative. The link between this simple inequality and the more abstract Peres-Horodecki criterion is recalled in Appendix~\ref{app:separability}. 

In Sec.~\ref{sec:system}, we studied the generic properties of the $S$-matrix on an analogue black hole horizon in order to adopt a parametrization of the spectra that takes into account the (super-luminal) dispersion relation. We then combined these parameters with those characterizing the three initial distributions of quasi-particles which are scattered on the horizon. The set of six parameters we used is described in Sec.~\ref{sec:sixparameters}. 

In Sec.~\ref{sec:domainsofnonsep}, we first studied the dependence in $\omega$ of $\Delta_\omega^{uu}$, the difference of \eq{eq:NSepuv} which characterizes the pairs of Hawking quanta. As expected, in the infra-red sector the state is separable, because stimulated effects dominate over spontaneous one. We also observed that the domain of non-separability critically depends on the strength of the couplings between the spectator third mode and the two modes under study. We then studied how the minimum value of $\Delta_\omega^{uu}$ over $\omega$ depends on the five parameters we adopted, see Eq.~\eqref{eq:sixparameters}. We showed that the domain of non-separability only depends on three combinations of these parameters. In addition, two of the three combinations possess two different forms. The parameter that distinguishes the two regimes is the ratio of the temperatures $T_H/T_{\infty}$, see Eq.~\eqref{Thor}. When it is smaller than $1/3$, dispersive effects are small, and one effectively works in a regime close to the relativistic Hawking one. Instead, when it is larger than $3$, one works in a regime where the surface gravity plays no significant role. We also showed that the crossover from one regime to the other is rather smooth. Combining the scalings in the two regimes, we obtained a rather complete characterization of the domains of non-separability, which we hope will be useful to guide future experiments to identify the appropriate range of parameters where the spontaneous Hawking effect dominates over stimulated ones. 

The main lesson is that in both regimes, the non-separability threshold critically depends on the initial temperature of the system (something which was expected), but also on the intensity of the coupling with the third spectator mode. When the latter is small enough, the final state can be quantum mechanically entangled even when the initial temperature is higher than the black hole temperature. For completeness, we also studied the quantity $\Delta_\omega^{uv}$ which governs the non-separability of the other type of pairs emitted by an analogue black hole. The relevant parameters are completely different.

In Appendix~\ref{app:sublum}, we briefly studied the modifications when replacing the superluminal dispersion relation by a subluminal one. No significant change is observed besides the fact that stimulated effects are slightly more important for subluminal dispersion, as the redshift of the initial distribution is less pronounced. In Appendix~\ref{app:whitehole}, we compared the entanglement obtained in a white hole flow with that of a black hole one. We showed that white holes are less appropriate to look for quantum entanglement because stimulated effects are much more important. 

\acknowledgments
We thank J.R.M. de Nova, F. Sols and I. Zapata for useful discussions during our visit in Madrid in November 2013. We also thank S. Finazzi, F. Michel, and S. Robertson for interesting remarks. This work has been supported by the French National Research Agency under the Program Investing in the Future Grant No. ANR-11-IDEX-0003-02 associated with the project QEAGE (Quantum Effects in Analogue Gravity Experiments).

\appendix

\section{Non-separability} 
\label{app:separability}

We consider two couples  of canonical conjugated operators, $(q_1,p_1)$ and $(q_2,p_2)$. The state of the two mode system spanned by these variables, $\hat \rho_2$, is obtained by tracing out all the other degrees of freedom. It can be characterized by the expectation values of these operators, their covariance matrix, and higher order polynomial. The Peres-Horodecki criterion~\cite{Peres:1996dw,Horodecki:1997vt,Simon:2000zz} is based on three $2 \times 2$ covariance matrices 
\begin{equation}
\begin{split}
A_i &= \Tr\left ( \rho \{X_i, X_i\}\right ),\quad C =  \Tr\left ( \rho \{X_1, X_2\}\right ), 
\end{split}
\end{equation}
where $X_i = [q_i -\Tr\left ( \rho q_i\right ) , p_i-\Tr\left ( \rho p_i\right )]$ is a two component vector, and where $i = 1,2$. The two relevant quantities are 
\begin{equation}
\label{eq:defPHcrit}
\begin{split}
\mathcal{P}_\pm \doteq &\det A_1 \, \det A_2 +(1/4\pm \abs{\det C})^2 \\
&- tr\left ( A_1 J C J A_2 J C^T J  \right ) - \left ( \det A_1 + \det A_2 \right )/4 , 
\end{split}
\end{equation}
where $J$ is the $2\times 2$ symplectic matrix:
\begin{equation}
\begin{split}
J = \left (\begin{array}{rr}
0 &1 \\
-1& 0 
\end{array}\right ).
\end{split}
\end{equation}
The quantity  $\mathcal{P}_+$ is always positive for physical states, see \cite{Simon:2000zz}. $\mathcal{P}_-$ corresponds to $\mathcal{P}_+$ for the state obtained by having performed a partial transpose ($p_1\to -p_1$, all other quantities fixed). The interest of $\mathcal{P}_-$ resides in the fact that when it is negative, it implies 
that the state $\hat \rho_2$ is non-separable in the sense of Werner~\cite{Werner:1989,Vidal:2002zz}. We recall that, by definition, bipartite separable states can be decomposed as 
\begin{equation}
\label{eq:defsepW}
\begin{split} 
\hat \rho_2  = \sum_a p_a \hat \rho_a^{(1)} \otimes \hat \rho_a^{(2)}
\end{split}
\end{equation}
where $p_a$ are positive and can thus be interpreted as probabilities, and where $\hat \rho_a^{(i)}$ are density matrices of one-mode sub-systems. Note that there exist non-separable states with $\mathcal{P}_- > 0$. However, these states are necessarily non-Gaussian. 

To make contact with the body of the paper, we introduce the destruction operators $a_i= (q_i + i p_i)/\sqrt{2}$, and their hermitian conjugated $a^\dagger_i$. We can then express the matrix elements of $A_i$ and $C$ in terms of occupation numbers and the coherence coefficients 
\begin{equation}
\begin{split}
n_{ij} \doteq \Tr\left ( \rho a_i^\dagger a_j\right ), \quad c_{ij} \doteq \Tr\left ( \rho a_i a_j\right ).
\end{split}
\end{equation}
In terms of those, one obtains 
\begin{equation}
\label{eq:valuePHcrit}
\begin{split}
\mathcal{P}_- =&   \left (n_{22} n_{11}  -\abs{c_{12}}^2 \right)\left (\left( 1+n_{22} \right)  \left( 1+n_{11} \right) -\abs{c_{12}}^2 \right) \\
& -\abs{n_{12}}^2 (n_{11}+ n_{22} + n_{11}n_{22}+ 2\abs{c_{12}}^2) \\
&-2\Re[c_{12}^*c_{11}n_{12}](1+2 n_{22})-2\Re[{c_{12}^*}^{2}c_{11}c_{22}]\\
&-2\Re[n_{12}c_{12}c_{22}^*](1+2 n_{11}) + \abs{{n_{12}}^{2} -c_{11}^*c_{22}}^2 \\ 
& - \abs{c_{22}}^2 n_{11}(1+n_{11})-\abs{c_{11}}^2 n_{22}(1+n_{22}) .
\end{split}
\end{equation}
When the state is stationary, and when $a_1, a_2$ are destruction operators with opposite frequency, the above expression simplifies a lot since $c_{ii} = n_{12} = 0$. In fact, $\mathcal{P}_-$ reduces to the first line. 
Then, since the second factor is positive, we have
\begin{equation}
\label{eq:sepequivalence}
\begin{split}
\mathcal{P}_-  < 0 \Leftrightarrow n_{22} n_{11}  < \abs{c_{12}}^2.
\end{split}
\end{equation}
Hence the negativity of $\mathcal{P}_- $ is equivalent to our non-separability condition $\Delta_{12} < 0$, see Eq.~\eqref{eq:NSepuv}. QED. 

When the two destruction operators $a_1, a_2$ have the same frequency $\omega>0$, all coefficients $c_{ij}$ vanish for stationary states. Hence we get
\begin{equation}
\label{eq:valuePHcritcis0}
\begin{split}
\mathcal{P}_-= &  (n_{22} n_{11}- \abs{n_{12}}^2) (n_{11}+ n_{22} + n_{11}n_{22})\\
&+ \abs{n_{12} }^4 + n_{22}^2 n_{11}^2 .
\end{split}
\end{equation}
This is positive because $n_{22} n_{11}- \abs{n_{12}}^2 \geq 0$. As a result, in agreement with Ref.~\cite{deNova:2012hm}, we conclude that the correlations between the $u$ and the $v$ modes with positive frequency, which are governed by the coefficient $d$ of Eq.~\eqref{eq:noutcoutdef}, cannot produce a non-separable (Gaussian) state. 

So far we used both Gaussianity and stationarity of the state to show that the positivity of $\cal P_-$ is {\it equivalent} to the positivity of $\Delta_{ij}$ for sectors with opposite frequency. However, when the state is non stationary, $\Delta_{ij} < 0$ still implies that the state is non separable. A rigorous proof can be found in Appendix B of Ref.~\cite{Adamek:2013vw}. To summarize, our criterion $\Delta_{ij} < 0$ is {\it sufficient} to guarantee the non-separability of the state. It also gives a {\it necessary} condition when the state is Gaussian and stationary.

\begin{figure}
\includegraphics[width= 1 \linewidth]{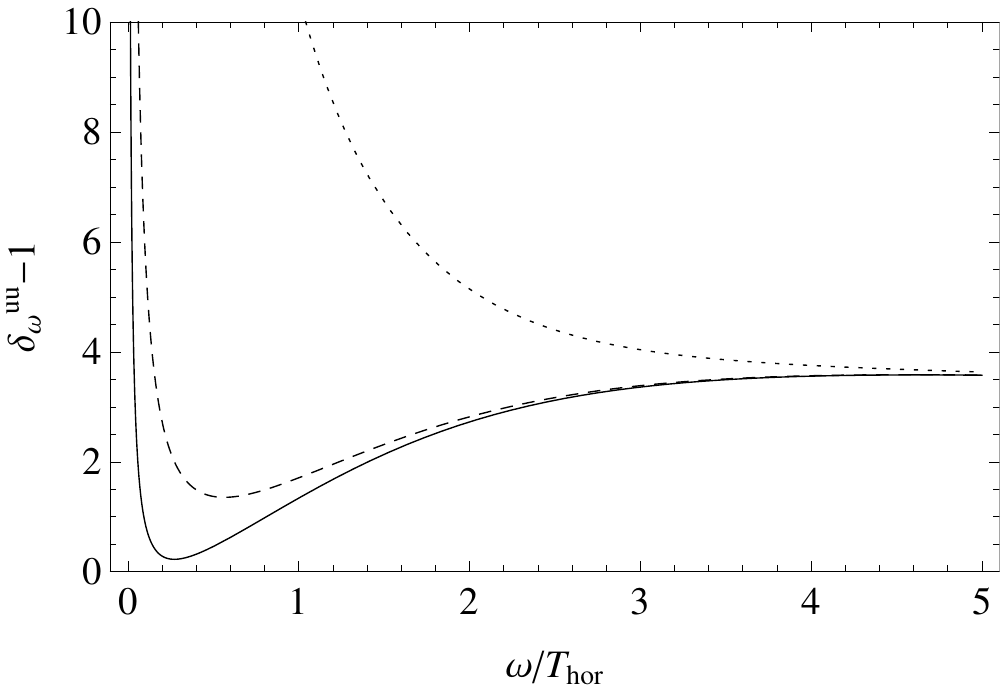}
\caption{The relative quantity $\delta_\omega^{uu}$ for the subluminal dispersion relation of \eq{drsub} as a function of $\omega/T_{\rm hor}$, for a high initial temperature $T_{\rm in} = 2  \Lambda$, for three values of $B= 0.01$ (solid), $0.02$ (dash), and $0.09$ (dot), $A = 4 B$, $\Lambda = 10 \kappa$, and $D=1/2$.} 
\label{fig:deltauofomegalargetempsub}
\end{figure}

\section{Subluminal dispersion relation}
\label{app:sublum}

We briefly consider the sub-luminal dispersion relation, 
\begin{equation}
\begin{split}
\label{drsub}
F^2(k) &= c^2 ( k^2 - \frac{k^4}{\Lambda^2}),
\end{split}
\end{equation}
in order to present the main differences with the non-separability of the super-luminal case considered in the body of the paper. In Fig.~\ref{fig:deltauofomegalargetempsub}, as in the right panel of Fig.~\ref{fig:deltauofomegalargetemp}, we represent the relative quantity $\delta_\omega^{uu}$ for a high initial temperature. For such temperature, we see that the state is slightly less entangled than in the superluminal case. The origin of this is due to the fact that $in$ modes come from the sub-sonic side of the horizon. As a result, for a given initial temperature $T_{\rm in}$, the effective $u$-temperature $T_{\rm in}^u$ of Eq.~\eqref{eq:ninofmuT} is larger than that found when the $in$ modes come from the supersonic side. In other words the initial distribution of $u$-quanta is less red-shifted for sub than super-luminal dispersion. This implies that the contribution of stimulated emission is higher, and this reduces the domains of non-separability.

The low temperature behavior of $\delta_\omega^{uu}$ is much less sensitive to the sign of the dispersion relation because in that case, the non-separability threshold is mainly governed by the coupling of the $v$ modes. Because there is no novel aspect in this case, we do not represent it. In addition, similar effects are also observed concerning the non-separability of $uv$ pairs. Hence, these need not to be studied separately.

\begin{figure}
\includegraphics[width= 1 \linewidth]{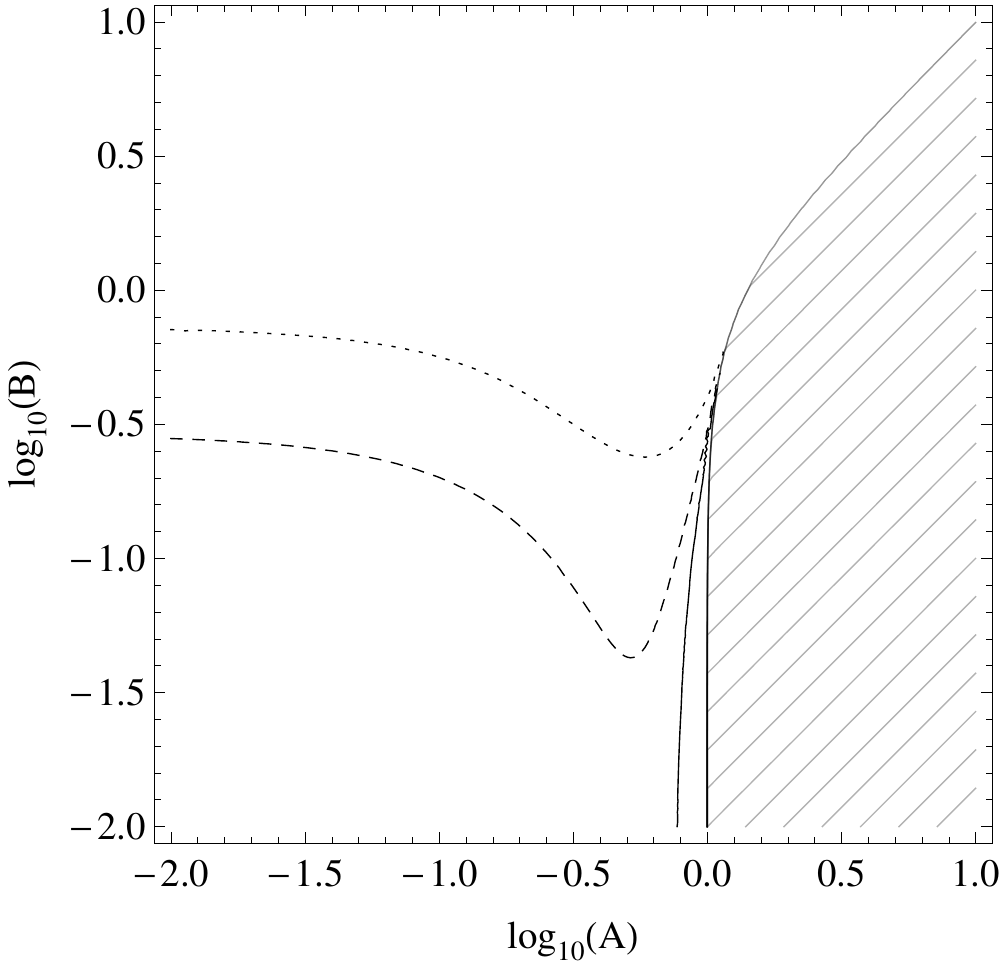}
\caption{The limit of non separability for the limit $\omega \to 0$ as a function of $A,B$. Parameters are $D= 1/2 $ and $T_{\rm in, WH}^{u} = 2 T_{\rm hor}  \times 0.5$(dotted), $0.8$ (dashed) and $1$ (solid).}
\label{fig:lowomegauunsep}
\end{figure}

\begin{figure*}
\begin{minipage}[t]{0.45\linewidth}
\includegraphics[width= 1 \linewidth]{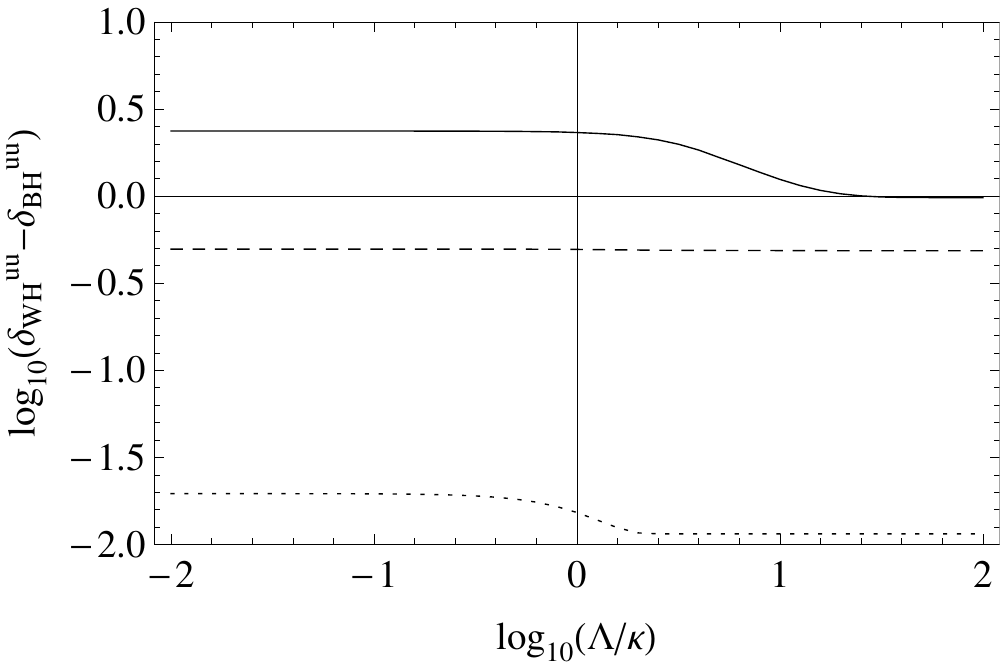}
\end{minipage}
\hspace{0.05\linewidth}
\begin{minipage}[t]{0.45\linewidth}
\includegraphics[width= 1 \linewidth]{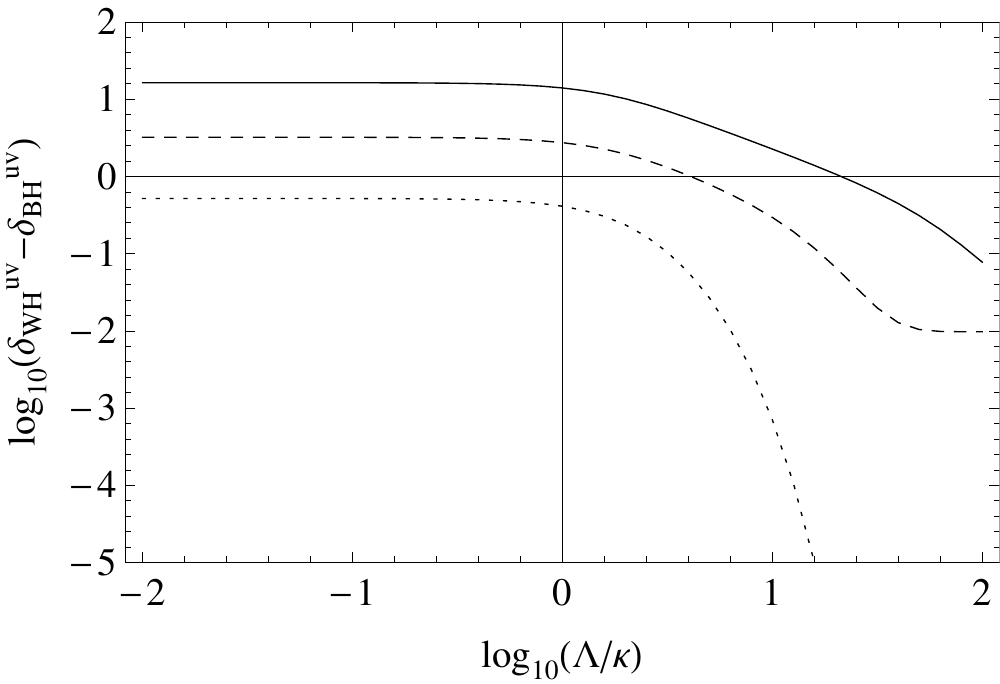}
\end{minipage}
\caption{The difference of the minima over $\omega$ of $\delta_{\omega}^{uu}$ (left panel) and $\delta_{\omega}^{uv}$ (right panel) computed in a white hole and in a black hole, for three different temperatures $T_{\rm in } = 2 T_{\rm hor} \times 0.2$ (dotted), $1$ (dashed), $5$ (solid) The parameters are $D=1/2, A=B =0.1$. One sees that increasing the initial temperature further increases the difference between the values of $\delta_{\omega}^{uu}$ and $\delta_{\omega}^{uv}$, which means that the non-separable character of the state is more rapidly lost for white holes than black holes.}
\label{fig:deltauuBHminusWH}
\end{figure*}

\section{Analogue white holes}
\label{app:whitehole}

We consider the white hole flow obtained by replacing $v(x)$ by $-v(x)$, where the flow profile $v(x)$ describes a black hole, for an example see Eq.~\eqref{eq:vpluscBH}. In this case, as explained in~\cite{Macher:2009tw}, $S_{WH}$, the $S$-matrix in the white hole flow is simply given by the inverse of the corresponding black hole one given in Eq.~\eqref{Bogcoef}. Because of unitarity, $S_{WH}$ is of the form $S_{WH} = T S^\dagger T$, where $T= {\rm diag}(1,-1,1)$. In terms of the black hole coefficients, $S_{WH}$ reads
\begin{equation}
S_{WH} = \left (\begin{array}{llll}
\alpha_\omega^* &- \beta_{-\omega}^* &\tilde A_\omega^* \\
 - \beta_\omega & \alpha_{-\omega}  &- \tilde B_\omega \\
A_\omega^*      &- B_\omega^*        & (\alpha^v_\omega)^* 
\end{array} \right ).
\end{equation}

When considering an incoherent initial state characterized by three initial occupation numbers,
Eq.~\eqref{eq:Sepsimplified} becomes
\begin{equation}
\begin{split}
\label{D-WH}
\Delta_\omega^{uu} &= \abs{\alpha_\omega^v}^2 n_{ \omega, {\rm WH}}^{u,\rm in}  n_{ -\omega, {\rm WH}}^{u,\rm in}  + \abs{ B_\omega}^2 n_{ \omega, {\rm WH}}^{u,\rm in}  n_{\omega, {\rm WH}}^{v, \rm in} \\
+& \abs{ A_\omega}^2 n_{ -\omega, {\rm WH}}^{u,\rm in}  n_{\omega, {\rm WH}}^{v, \rm in} + \abs{\tilde B_\omega}^2 n_{ \omega, {\rm WH}}^{u,\rm in}  +  \abs{\beta_{\omega}}^2 n_{\omega, {\rm WH}}^{v, \rm in} \\
-& \abs{\beta_{-\omega}}^2 (1+ n_{\omega, {\rm WH}}^{v, \rm in} + n_{ \omega, {\rm WH}}^{u,\rm in}  + n_{ -\omega, {\rm WH}}^{u,\rm in} )\\
\Delta_\omega^{uv} &= \abs{\tilde A_\omega}^2 n_{ \omega, {\rm WH}}^{u,\rm in}  n_{ -\omega, {\rm WH}}^{u,\rm in}  + \abs{\beta_{-\omega}}^2 n_{ \omega, {\rm WH}}^{u,\rm in}  n_{\omega, {\rm WH}}^{v, \rm in} \\
+& \abs{\alpha_\omega}^2 n_{ -\omega, {\rm WH}}^{u,\rm in}  n_{\omega, {\rm WH}}^{v, \rm in} + \abs{\tilde B_\omega}^2 n_{ \omega, {\rm WH}}^{u,\rm in}  +  \abs{\beta_{\omega}}^2 n_{\omega, {\rm WH}}^{v, \rm in} \\
-& \abs{ B_\omega}^2 (1+ n_{\omega, {\rm WH}}^{v, \rm in} + n_{ \omega, {\rm WH}}^{u,\rm in}  + n_{ -\omega, {\rm WH}}^{u,\rm in} ).
\end{split}
\end{equation} 
On the other hand, working again with the thermal initial state of \eq{eq:ninofOmega}, 
in the place of \eq{eq:ninofmuT}, the initial distributions are
\begin{equation}
\begin{split}
n_{\omega, {\rm WH}}^{v, \rm in} \sim \frac{1}{\exp\left (  \omega / T_{\rm in, WH}^v \right ) -1}, \\ 
n_{ \pm \omega, {\rm WH}}^{u,\rm in} \sim \frac{1}{\exp\left (  \omega /T_{\rm in, WH}^{u} \right ) -1} ,
\end{split}
\end{equation}
where 
\begin{equation}
\begin{split}
T_{\rm in, WH}^{u} =  T_{\rm in} D, \quad 
T_{\rm in, WH}^v = T_{\rm in} (2  +  D) . 
\end{split}
\end{equation}
These two effective temperatures are independent of the dispersion relation because the three incoming modes are now low momentum ones. As a result, for white hole flows, the dispersive scale $\Lambda$ only enters in the final distributions only through the effective temperature $T_{\rm hor}$ of Eq.~\eqref{Thor}, which is the same for the black and the white hole flows $\pm v(x)$.

When considering the entanglement of the quasi-particles emitted by a white hole, one expects that it will be weaker than that of the corresponding black hole. The reason is clear: in white hole flows, stimulated effects dominate in over the spontaneous channel because low frequency excitations are blue shifted (at fixed $\omega$, the final value of the wave number $k_\omega$ is larger than the incoming one). As a result, 
the quantities of \eq{D-WH} diverge in the low frequency limit as
\begin{equation}
\begin{split}
\omega^2 \times \Delta_\omega^{uu} &\underset{\omega\to 0} \sim   \alpha_v^2 (T_{\rm in, WH}^{u})^2 + (A^2 + B^2) T_{\rm in, WH}^{u} T_{\rm in, WH}^v \\
&- T_{\rm hor} \left [ ( \gamma_- - \gamma_+ ) T_{\rm in, WH}^v  + (\gamma_- + \gamma_+ )T_{\rm in, WH}^{u}  \right ]\\
\omega^3 \times \Delta_\omega^{uv} &\underset{\omega\to 0}\sim T_{\rm in, WH}^{u} T_{\rm hor} \times \\
& \left [ ( \gamma_- - \gamma_+ ) T_{\rm in, WH}^{u} + (\gamma_-+\gamma_+)  T_{\rm in, WH}^v \right ] , \\
\end{split}
\end{equation}
where $\gamma_\pm$ are the two quantities defined after Eq.~\eqref{eq:Deltaulowomega}. 

The main consequence of these equations is that the non-separability can be effectively studied by considering the low frequency limit. In Fig.~\ref{fig:lowomegauunsep}, we represent the limit of non separability $\Delta_{\omega}^{uu} = 0$ at $\omega = 0$ in the $(A ,B)$ plane. For $T_{\rm in, WH}^{u} < 2 T_{\rm hor}$, we observe that the state is non-separable in a very large domain of the plane. Instead, for $T_{\rm in, WH}^{u} \geq 2 T_{\rm hor}$, only a small domain remains non separable. The transition between the two regimes is rapid since changing the value of the temperature by $20\%$ is sufficient to obtain non separability for $B \lesssim 0.3$. We verified that these conclusions are not significantly modified by relaxing the condition $\omega \to 0$, and taking the minimum of $\delta_{\omega}^{uu}$ as done in the body of the text. For $T_{\rm in, WH}^{u} < 2 T_{\rm hor}$, we observed an increase of the non separability domain. Instead for $T_{\rm in, WH}^{u} \geq 2 T_{\rm hor}$, we did not observe any increase. 

To conclude this Appendix, it is interesting to compare the values of $\delta_{\omega}^{uu}$ computed in a white hole and in a black hole for the same initial state. In the left panel of Fig.~\ref{fig:deltauuBHminusWH} (resp. the right panel) we represent the difference of the minima over $\omega$ of $\delta_\omega^{uu}$ (resp. $\delta_\omega^{uv}$) of Eq.~\eqref{eq:Nsepuvrelat} between the black hole and the white hole case. We used the sames values of the parameters $D=1/2, A=B =0.1$, $T_{\rm in}/T_{\rm hor}$ and plot the dependence of the function in $\Lambda/ \kappa$. We observe that $\delta$ is generically higher in the white hole flow than in the corresponding black hole one.

\bibliographystyle{../biblio/h-physrev}
\bibliography{../biblio/bibliopubli} 

\begin{thebibliography}{10}

\bibitem{Unruh:1980cg}
W.~Unruh,
\newblock Phys.Rev.Lett. {\bf 46}, 1351 (1981).

\bibitem{Barcelo:2005fc}
C.~Barcelo, S.~Liberati, and M.~Visser,
\newblock Living Rev. Rel. {\bf 14}, 3 (2011), gr-qc/0505065.

\bibitem{Schutzhold:2002rf}
R.~Schutzhold and W.~G. Unruh,
\newblock Phys.Rev. {\bf D66}, 044019 (2002), gr-qc/0205099.

\bibitem{Rousseaux:2007is}
G.~Rousseaux, C.~Mathis, P.~Maissa, T.~G. Philbin, and U.~Leonhardt,
\newblock New J.Phys. {\bf 10}, 053015 (2008), 0711.4767.

\bibitem{Weinfurtner:2010nu}
S.~Weinfurtner, E.~W. Tedford, M.~C. Penrice, W.~G. Unruh, and G.~A. Lawrence,
\newblock Phys.Rev.Lett. {\bf 106}, 021302 (2011), 1008.1911.

\bibitem{Unruh:1994je}
W.~Unruh,
\newblock Phys.Rev. {\bf D51}, 2827 (1995).

\bibitem{Brout:1995wp}
R.~Brout, S.~Massar, R.~Parentani, and P.~Spindel,
\newblock Phys.Rev. {\bf D52}, 4559 (1995), hep-th/9506121.

\bibitem{Carusotto:2008ep}
I.~Carusotto, S.~Fagnocchi, A.~Recati, R.~Balbinot, and A.~Fabbri,
\newblock New J.Phys. {\bf 10}, 103001 (2008), 0803.0507.

\bibitem{Macher:2009nz}
J.~Macher and R.~Parentani,
\newblock Phys. Rev. {\bf A80}, 043601 (2009), 0905.3634.

\bibitem{Recati:2009ya}
A.~Recati, N.~Pavloff, and I.~Carusotto,
\newblock Phys.Rev. {\bf A80}, 043603 (2009), 0907.4305.

\bibitem{Bruschi:2013tza}
D.~E. Bruschi, N.~Friis, I.~Fuentes, and S.~Weinfurtner,
\newblock New J. Phys. 15, {\bf 113016} (2013), 1305.3867.

\bibitem{Busch:2013sma}
X.~Busch and R.~Parentani,
\newblock Phys.Rev. {\bf D88}, 045023 (2013), 1305.6841.

\bibitem{Busch:2013gna}
X.~Busch, I.~Carusotto, and R.~Parentani,
\newblock (2013), 1311.3507.

\bibitem{PhysRevLett.108.260401}
K.~V. Kheruntsyan {\em et~al.},
\newblock Phys. Rev. Lett. {\bf 108}, 260401 (2012).

\bibitem{WilsonDCE}
C.~M. Wilson {\em et~al.},
\newblock Nature {\bf 479}, 376–379 (2011), 1105.4714.

\bibitem{Campo:2005sy}
D.~Campo and R.~Parentani,
\newblock Phys.Rev. {\bf D72}, 045015 (2005), astro-ph/0505379.

\bibitem{Campo:2005sv}
D.~Campo and R.~Parentani,
\newblock Phys.Rev. {\bf D74}, 025001 (2006), astro-ph/0505376.

\bibitem{Campo:2008ju}
D.~Campo and R.~Parentani,
\newblock Phys.Rev. {\bf D78}, 065044 (2008), 0805.0548.

\bibitem{deNova:2012hm}
J.~de~Nova, F.~Sols, and I.~Zapata,
\newblock (2012), 1211.1761.

\bibitem{finazzi2013}
S.~{Finazzi} and I.~{Carusotto},
\newblock ArXiv e-prints  (2013), 1309.3414.

\bibitem{Finazzi:2012iu}
S.~Finazzi and R.~Parentani,
\newblock Phys.Rev. {\bf D85}, 124027 (2012), 1202.6015.

\bibitem{Macher:2009tw}
J.~Macher and R.~Parentani,
\newblock Phys.Rev. {\bf D79}, 124008 (2009), 0903.2224.

\bibitem{Robertson:2011xp}
S.~J. Robertson,
\newblock (2011), 1106.1805.

\bibitem{Coutant:2011in}
A.~Coutant, R.~Parentani, and S.~Finazzi,
\newblock Phys.Rev. {\bf D85}, 024021 (2012), 1108.1821.

\bibitem{Robertson:2012ku}
S.~J. Robertson,
\newblock J.Phys. {\bf B45}, 163001 (2012).

\bibitem{Zapata:2011ze}
I.~Zapata, M.~Albert, R.~Parentani, and F.~Sols,
\newblock New J.Phys. {\bf 13}, 063048 (2011), 1103.2994.

\bibitem{Coutant:2009cu}
A.~Coutant and R.~Parentani,
\newblock Phys.Rev. {\bf D81}, 084042 (2010), 0912.2755.

\bibitem{Finazzi:2010nc}
S.~Finazzi and R.~Parentani,
\newblock New J.Phys. {\bf 12}, 095015 (2010), 1005.4024.

\bibitem{Finazzi:2010yq}
S.~Finazzi and R.~Parentani,
\newblock Phys.Rev. {\bf D83}, 084010 (2011), 1012.1556.

\bibitem{Adamek:2013vw}
J.~Adamek, X.~Busch, and R.~Parentani,
\newblock Phys. Rev. {\bf D 87} (2013), 1301.3011.

\bibitem{Peres:1996dw}
A.~Peres,
\newblock Phys.Rev.Lett. {\bf 77}, 1413 (1996), quant-ph/9604005.

\bibitem{Horodecki:1997vt}
P.~Horodecki,
\newblock Phys.Lett. {\bf A232}, 333 (1997), quant-ph/9703004.

\bibitem{Simon:2000zz}
R.~Simon,
\newblock Phys.Rev.Lett. {\bf 84}, 2726 (2000).

\bibitem{Werner:1989}
R.~F. Werner,
\newblock Phys. Rev. A {\bf 40}, 4277 (1989).

\bibitem{Vidal:2002zz}
G.~Vidal and R.~Werner,
\newblock Phys.Rev. {\bf A65}, 032314 (2002).

\end{thebibliography}

\end{document}